\documentclass[useAMS,usenatbib]{mn2e}
\usepackage{graphicx}
\usepackage{amsmath}
\usepackage{fixltx2e} 
\usepackage[backref,breaklinks,colorlinks,citecolor=blue]{hyperref}
\usepackage[all]{hypcap} 

\title[The Polar LSQ1725-64]{The Magnetic Cataclysmic Variable LSQ1725-64\thanks{Based on observations obtained at the Southern Astrophysical Research (SOAR) telescope, which is a joint project of the Minist\'{e}rio da Ci\^{e}ncia, Tecnologia, e Inova\c{c}\~{a}o (MCTI) da Rep\'{u}blica Federativa do Brasil, the U.S. National Optical Astronomy Observatory (NOAO), the University of North Carolina at Chapel Hill (UNC), and Michigan State University (MSU). Also based on observations made with the Southern African Large Telescope (SALT).}}

\author[J.T. Fuchs et al.]{J.T.~Fuchs,$^1$\thanks{joshfuchs@unc.edu} Bart H.~Dunlap,$^1$ E.~Dennihy,$^1$ D.~O'Donoghue,$^{2,3}$ J.C.~Clemens,$^1$ 
\newauthor 
D.E.~Reichart,$^1$ J.P.~Moore,$^1$ A.P.~LaCluyze,$^1$ J.B.~Haislip,$^1$ and K.V.~Ivarsen,$^1$\\
  $^1$ Department of Physics and Astronomy, University of North Carolina at Chapel Hill, Chapel Hill, NC 27599, USA\\
$^2$ South African Astronomical Observatory, PO Box 9, Observatory 7935, South Africa\\
$^3$ Southern African Large Telescope, PO Box 9, Observatory 7935, South Africa}
\date{Accepted 2016 July 18. Received 2016 July 18; in original form 2016 March 22}
\begin{document}

\maketitle

\begin{abstract}
We present new photometry and spectroscopy of the 94m eclipsing binary LSQ1725-64 that provide insight into the fundamental parameters and evolutionary state of this system. We confirm that LSQ1725-64 is a magnetic cataclysmic variable whose white dwarf has a surface-averaged magnetic field strength of $12.5 \pm 0.5$ MG measured from Zeeman splitting. The spectral type and colour of the secondary, as well as the eclipse length, are consistent with other secondaries that have not yet evolved through the period minimum expected for cataclysmic variables. We observe two different states of mass transfer and measure the transition between the two to occur over about 45 orbital cycles. In the low state, we observe photometric variations that we hypothesize to arise predominantly from two previously heated magnetic poles of the white dwarf. Our precise eclipse measurements allow us to determine binary parameters of LSQ1725-64 and we find it contains a high mass ($0.97 \pm 0.03\ M_{\odot}$) white dwarf if we assume a typical mass-radius relationship for a CO core white dwarf. We also measure an eclipse of the accretion stream after the white dwarf eclipse, and use it to estimate an upper limit of the mass transfer rate. This derived limit is consistent with that expected from angular momentum loss via gravitational radiation alone. 
\end{abstract}

\begin{keywords}
stars: cataclysmic variables, binaries: eclipsing, stars: individual: LSQ172554.8-643839, magnetic fields, accretion
\end{keywords}
\section{Introduction}
Most cataclysmic variables are white dwarf - main sequence binaries in orbits so close that the main sequence companion sporadically or continuously transfers matter to the white dwarf. \citet{Warner:1995CASa} provides a thorough summary of the field. Around 25 per cent of the white dwarfs in cataclysmic variables have magnetic field strengths of at least a few MG \citep{Wickramasinghe:2000Pa}.

Polars, or AM Her stars, are an observationally distinguished subclass that contain white dwarfs with magnetic fields greater than about 10 MG. They exhibit strong (greater than ten per cent) linear and circular polarization \citep{Cropper:1990a}. Their spectra show strong Balmer emission lines along with He I and He II. He II is typically stronger in polars compared to non-magnetic cataclysmic variables \citep{Szkody:1998a}. Cyclotron radiation from spiraling matter in the accretion stream can be seen in time-resolved spectroscopy. The wavelength of the cyclotron harmonics depends on the magnetic field strength of the white dwarf and is commonly in the infra-red or optical \citep{Cropper:1989Ma}. Measuring the Zeeman splitting of the Balmer lines can give an estimate of the white dwarf magnetic field strength and geometry \citep{Beuermann:2007a}.

Polars tend to have lower rates of mass transfer, around $10^{-11} M_{\odot}\  yr^{-1}$, compared to non-magnetic cataclysmic variables \citep{Patterson:1984Aa}. On time-scales that are not well constrained, the mass transfer rate can drop by an order of magnitude or more (see \citet{Warner:1999a} and \citet{Hessman:2000a}). These low states provide an opportunity to study the stellar components more easily than the high states, when light from accretion can swamp measurements of the stellar photospheres. Low states typically last between days and months, though there is increasing evidence that some polars stay in the low states for years, e.g. EF Eri, which has mostly been in a low state since 1997 \citep{Wheatley:1998Aa} with only brief returns to the high state since 2006.

LSQ172554.8-643839 (hereafter LSQ1725-64) is an eclipsing binary with a period of $\sim$94 minutes. Its \textit{V}-band magnitude varies between 18 and \textgreater 24 throughout the orbital cycle. It was first discovered as a variable star by \citet{Rabinowitz:2011a}. Follow-up optical photometry showed an eclipse that lasted for $\sim$5 minutes with an ingress depth of more than 5.7 mag; amongst the deepest eclipses known for all cataclysmic variables. \citet{Rabinowitz:2011a} measured a magnitude for the secondary star during eclipse in the \textit{J} band only; without a colour measurement they were only able to place a limit on the spectral class. They did not measure any evidence for a magnetic field, so could not definitely conclude that LSQ1725-64 is a magnetic cataclysmic variable.

The properties of LSQ1725-64 as described by \citet{Rabinowitz:2011a} make it an intriguing and appealing object for follow-up study. It could offer insight into magnetic accretion physics, the role of magnetic fields in binary star evolution, and the period evolution of cataclysmic variables. This last item is intimately related to the mass-radius relation of the secondary star and the mechanisms that might drain angular momentum from binaries: magnetic braking and gravitational radiation. Short period systems like LSQ1725-64 are particularly interesting to study, as there should be some period at which continued mass transfer causes the secondary to expand above its equilibrium radius, resulting in evolution to longer rather than shorter periods. The location of this period bounce and the evolution of systems that have passed through it are critical to our understanding of angular momentum loss and period evolution in cataclysmic variables.

Based on derived binary parameters and a \textit{V-J} limit during eclipse, \citet{Rabinowitz:2011a} suggested LSQ1725-64 might be a post-bounce system, i.e. one that is already evolving to longer orbital periods as the now degenerate secondary expands in response to mass loss. This would make it a rare and valuable object of study as there are very few known (see e.g. \citet{Littlefair:2006Sa} and \citet{Gansicke:2009Ma}). On the basis of these interesting possibilities, we conducted a campaign of time-resolved spectroscopy, short cadence photometry, and photometric monitoring of LSQ1725-64.

In \autoref{sec:observations}, we describe our observations and reduction process. In \autoref{sec:bfield}, we show spectroscopic measurements of the H $\beta$ absorption line of the white dwarf that permit us to conclude with certainty that the system contains a magnetic white dwarf. In \autoref{sec:photometric}, we summarize the photometric behaviour of LSQ1725-64 on time-scales ranging from seconds to years. We present results from time-resolved optical spectroscopy in \autoref{sec:spectroscopic}. Section \ref{sec:discussion} discusses the evolutionary status and presents updated binary parameters of LSQ1725-64. Finally, we conclude in \autoref{sec:conclusions} by fitting LSQ1725-64 into context as a polar and adding our measurements to what is already known.

\section{Observations}
\label{sec:observations}
Our new observational data include three kinds of observations of LSQ1725-64: time-series spectroscopy, short cadence photometry, and photometric monitoring of accretion state. As summarized in \autoref{tab:goodmanphot} and \autoref{tab:goodmanspec}, we observed LSQ1725-64 on 14 separate nights with an imaging spectrograph---the Goodman Spectrograph on the 4.1-m SOAR Telescope \citep{Clemens:2004a}.

For the SOAR Goodman photometry, we binned the CCD at 2 $\times$ 2 with the exception of 2012-08-12, when we binned at 1 $\times$ 1. We did not use a filter for most of the photometry, allowing us to decrease the exposure times and increase the amount of time-dependent geometrical information in the data. For select eclipse measurements we obtained photometry in SDSS \textit{u}$'$, \textit{g}$'$, \textit{r}$'$, and \textit{i}$'$ bands. To reduce the readout time, the typical region of interest used was 90 $\times$ 90 arcsec, resulting in a readout time of $\sim$ 3 s and a duty cycle greater than 80 per cent. All observations beginning in 2013 have GPS derived shutter open times recorded in the image header. \autoref{tab:goodmanphot} gives details of all our photometry.

\begin{table*}
\begin{minipage}{130mm}
  \caption{Photometric Observations with the Goodman Spectrograph on the SOAR Telescope and SALTICAM on SALT.}
  \label{tab:goodmanphot}
\begin{tabular}{c c c c c c}
\hline
Observation Start & Instrument & Exposure & Length of & Eclipses & Accretion \\
Date (UT) & & Time (s) & Observation (hours) & Observed &State\\
\hline
2011-09-07 & SALTICAM & 15.7 & 1.8 & 1& High\\
2012-08-12 & Goodman & 20 & 4.5 & 3 & High\\
2013-06-08 & Goodman & 20 & 1.7 & 1.5 & Low\\
2013-07-03 & Goodman & 20 & 2.3 & 2 & High\\
2013-07-06 & SALTICAM & 1.7 & 0.8 & 1 & High\\
2013-07-12 & Goodman & 12 & 2.9 & 2 & High\\
2013-07-27 & Goodman & 12 & 1.6 & 1 & High\\
2013-08-05 & Goodman & 12 & 3.6 & 2 & High\\
2013-08-14 & Goodman & 11 & 2.0 & 2 & High\\
2013-08-15 & Goodman & 12 & 1.2 & 1 & High\\
2013-09-02 & SALTICAM & 1.7 & 0.8 & 1 & High\\
2013-11-14 & Goodman & 12 & 1.8 & 1 & High\\
2014-06-30 & Goodman & 12 & 2.5 & 2  & High\\
\hline
\end{tabular}
\end{minipage}
\end{table*}

For our spectroscopic observations, we used a 1.68$''$ slit, but did not stay aligned to the parallactic angle. There was no atmospheric dispersion corrector installed on the Goodman Spectrograph at the time of these observations. We took continuous sequences of spectra with no gaps, which limited the measurement of standards to only one a night. As a result of this observing strategy, our flux calibrations are less than ideal, but our velocity measurements, with errors $\sim 40\ km\ s^{-1}$, are as good as the instrument can do given our resolution.

Nearly all spectroscopic observations had exposure times of 300 s, equalling a phase width of 5.2 per cent. During eclipse, or the low state, the typical signal-to-noise of the continuum from an individual spectrum was around 6 per binned pixel. In the high state, the typical signal-to-noise of the continuum from an individual spectrum of the accreting hemisphere was between 8 and 13 per binned pixel. \autoref{tab:goodmanspec} gives details of all our spectroscopy.

\begin{table*}
\begin{minipage}{140mm}
  \caption{SOAR Goodman Spectroscopic Observations}
  \label{tab:goodmanspec}
\begin{tabular}{c c c c c c}
\hline
Observation Start & Grating  & Exposure & Resolution  & Length of & Accretion \\
 Date (UT) & (l mm$^{-1}$)& Time (s)& ($\AA$) &  Observation (hours)& State\\
\hline
2012-09-26 & 400 &  300 & 12.7 & 1.43 & High \\
2013-06-07 & 400 & 300 & 12.8 & 1.16 & Low \\
2013-06-08 & 400 & 300 & 15.3 & 1.68 & Low \\
2013-06-16 & 400 & 300 & 10.2 & 3.20 & High\\
2013-07-02 & 400 & 300 & 12.0 & 1.60 & High \\
2013-07-03 & 930 & 300, 600, 60 & 4.9 &  2.00 & High \\
2013-07-12 & 400 & 300 & 12.6 & 2.26 & High \\
2013-07-27 & 400 & 300 &  10.2 & 2.52 & High\\
\hline
\end{tabular}
\end{minipage}
\end{table*}

We also observed LSQ1725-64 three times with the SALTICAM Imager on the Southern African Large Telescope (SALT; \citet{ODonoghue:2006Ma}). We used frame transfer mode on SALTICAM, which allows half the chip to be shifted on to the unexposed half of the chip nearly instantaneously, limiting dead time between exposures. As with the Goodman photometry, we did not use a filter. The observations with SALT did not cover a full orbit, but instead were centred on the eclipse to resolve the ingress and egress structure. 

We used the PROMPT network of robotic telescopes to monitor the accretion state of LSQ1725-64. As with the other photometry, we did not use a filter. Exposure times were 180 s, with a readout time of $\sim$6 s, providing a duty cycle of 96.7 per cent. PROMPT monitored LSQ1725-64 on 70 separate nights during low, high, and intermediate states. We typically obtained one orbit per night.

We reduced and analyzed all photometry in the same way. The data were bias subtracted but not flat-fielded, because in unfiltered light the flat fields display large-scale, non-stable interference patterns. Lightcurves were produced using APER, an IDL function based on DAOPHOT \citep{Stetson:1987Pa}. All photometry makes use of the same three nearby comparison stars. To correct for long-term environmental trends in the lightcurve, we divided the flux from LSQ1725-64 by the average flux from the three comparison stars. We then define the mean flux as the average in region 2 on the night of 2013-07-02. All photometry is presented relative to this average.

Spectroscopic observations were bias subtracted but were also not flat-fielded. The spectra were extracted and wavelength calibrated using the standard \textsc{IRAF} tasks and a HgAr lamp. We produced rough flux calibrations by measuring the instrument response with a standard star. We did not sample the same range of airmasses in these standards as in the data, so the main sources of error in the flux calibration are changes in extinction and differential slit losses.

\section{The Magnetic Field of the White Dwarf}
\label{sec:bfield}
The observed low state in LSQ1725-64 allowed us to measure the Zeeman splitting of H $\beta$ in our spectroscopic observations. For magnetic fields less than $\sim$20 MG, the linear Zeeman effect produces a triplet pattern around the central absorption feature, calculable by 

\begin{equation}
\Delta \lambda_L \simeq 4.7 \times 10^{-7}\ \lambda^2\ B_s\ ,
\label{eq:zeeman}
\end{equation}
where $\lambda$ is in \AA, $B_s$ is the surface magnetic field in MG, and $\Delta \lambda_L$ is the separation of the split components from the central component \citep{Landstreet:1980a}. 

\begin{figure}
  \includegraphics[width=0.5\textwidth]{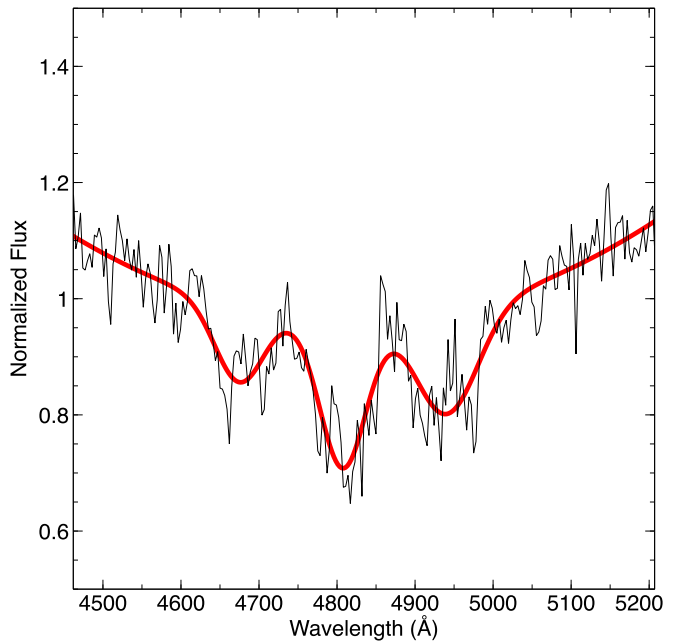}
  \caption{Orbital-averaged low state spectrum of LSQ1725-64 showing Zeeman splitting of the H $\beta$ line. The fit to determine the splitting is shown in red. The best-fitting model indicates a surface-averaged magnetic field of $12.5 \pm 0.5$ MG, consistent with white dwarf magnetic field strengths expected in polars.}
  \label{fig:zeeman}
\end{figure}

The H $\beta$ line has a high signal-to-noise and is relatively free from emission. \autoref{fig:zeeman} shows an orbital-averaged spectrum of LSQ1725-64 in the low state. We fitted the H $\beta$ line with three equally spaced gaussians to model the Zeeman absorption and used a second-order polynomial for the continuum. The result is shown in red. From \autoref{eq:zeeman} we determine a surface-averaged magnetic field of $12.5 \pm 0.5$ MG.

This magnetic field strength of the white dwarf falls within what is expected for polars. Our measurement of Zeeman splitting offers the best confirmation that the white dwarf in LSQ1725-64 has a strong magnetic field, and allows definitive classification of the binary as a cataclysmic variable of the AM Her type. 

\section{Photometric Properties of LSQ1725-64}
\label{sec:photometric}

In this section, we will present and discuss photometric data gathered on time-scales ranging from seconds to months, but first we update the orbital ephemeris of the binary.

\subsection{Orbital Ephemeris}
\label{sec:ephemeris}
\begin{figure}
  \includegraphics[width=0.5\textwidth]{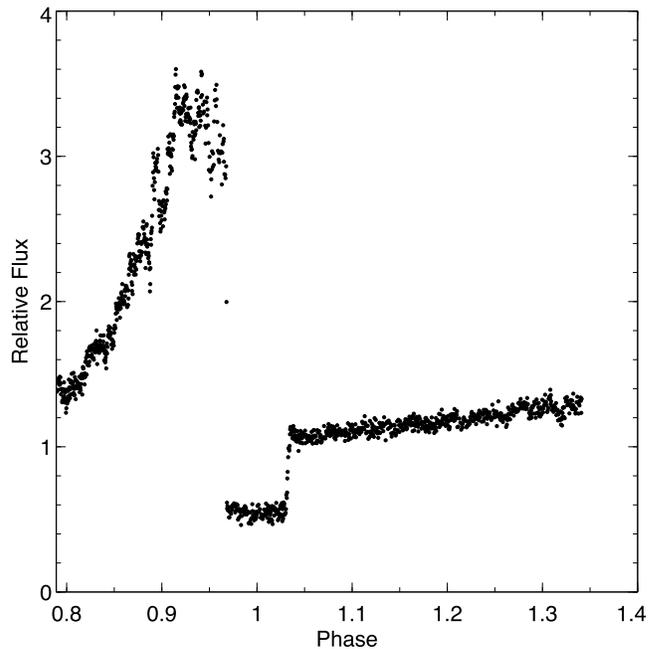}
  \caption{A SALT lightcurve from 2013-07-06 with an exposure time of 1.7 s. The structure of the eclipse ingress of the white dwarf and hotspot is not resolved with this time sampling. The egress of the white dwarf lasts 23.4 $\pm$ 0.3 s and appears unaffected by any hotspot.}
  \label{fig:salt}
\end{figure}

The eclipses in this system yield a very precise measurement of the orbital period. The SALT lightcurve shown in \autoref{fig:salt} displays a single eclipse at 1.7 s time resolution, taken when the polar was in a high state. In this state, the eclipse ingress is preceded by flickering and includes both the eclipse of the white dwarf and of the accreting pole, which is very near to the limb. This makes it difficult to measure the time of mid-ingress. The eclipse egress does not show this behaviour and lasts 23.4 $\pm$ 0.3 s. Thus, we measured eclipse times for all of our photometric data using times of mid-egress. The measured mid-egress times were all adjusted to Barycentric Julian Date in the Barycentric Dynamical Time standard ($BJD_{TDB}$) using the code of \citet{Eastman:2010Pa}. Only SOAR and SALT eclipse data were used to calculate the ephemeris.

To estimate mid-egress times from our data, we fitted a piece-wise function to the egress of the white dwarf. We modeled the eclipse as a linear change between constant flux levels before and after egress. The fit was done with MPFIT, which uses a Levenberg-Marquardt technique to solve the least-squares problem \citep{Markwardt:2009a}. The linear model was integrated over the same exposure times as the data during the fitting process. All fitted mid-egress times and their associated errors are shown in Appendix \ref{app:timings}.

We constructed an O-C diagram using the mid-egress times of the white dwarf. $T_0$ was chosen to be the SALT data point taken on 2013-07-06 as it had small error bars and was near the midpoint of all data presented here. We show the O-C diagram in Online Figure 1. Within the precision of these data, we are not able to detect any changes in the orbital period, as might occur either from angular momentum losses or from reflex motion generated by a third body. Therefore, we iteratively fitted a linear ephemeris to get the following result:

\begin{equation} \label{eq:ephemeris}
\begin{split}
T_0 =\  &2456480.4422112(55)\ +  \\
  & E\ \times\ 0.065741721(2)\ days\ (BJD_{TDB})
\end{split}
\end{equation}
where the numbers in parentheses are the error on the last digits. This period corresponds to 5680.0847(1) s. Note that the $T_0$ is a time of mid-egress and throughout this paper we define orbital phase 0 as the mid-eclipse time. Thus, $T_0$, the time of mid-egress, occurs at orbital phase 0.033.  This phase difference was established by measuring the time from end of ingress to beginning of egress, and combining half of that value with half of the egress time. With this ephemeris in hand, we can proceed to look at data over a single orbit or folded over several cycles.

\subsection{Photometric Variations at the Orbital Period}
\label{sec:nomenclature}	

\begin{figure*}
\centering
\begin{minipage}[c]{\textwidth}
\centering
    \includegraphics{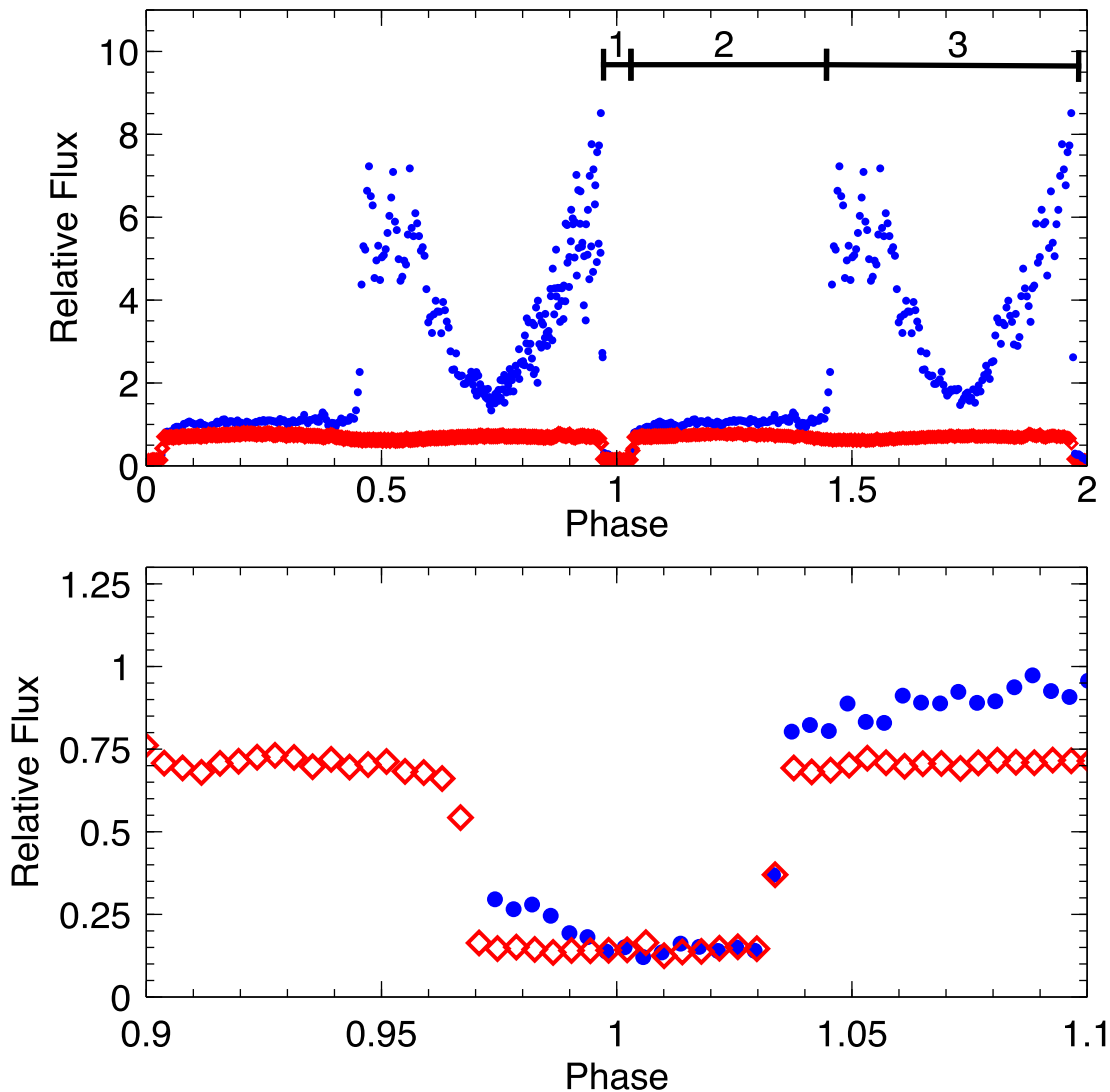}
  \caption{A comparison of the two states of mass transfer in LSQ1725-64. Blue circles show the high state from 2013-07-02. Red diamonds show the low state from 2013-06-08. The bottom plot shows the same data, but enlarged to show the slow eclipse of the accretion stream above the limb of the white dwarf after the eclipse of the white dwarf and hotspot. The high state data are above the scale shown. We mark regions 1, 2, and 3 as defined by \citet{Rabinowitz:2011a} at the top of the upper panel. We present the photometry as a flux relative to the average of region 2.}
  \label{fig:HighLowPhot}
\end{minipage}
\end{figure*}

The two lightcurves shown in \autoref{fig:HighLowPhot} exemplify the white light photometric behaviour of LSQ1725-64 over its orbital cycle. In this section we will describe how the variations in both the high and low state translate into geometrical information about LSQ1725-64.

\subsubsection{ Photometric Variations During the High State}
\label{sec:phot_variations_high}

As \autoref{fig:HighLowPhot} shows, our photometry of LSQ1725-64 exhibits the same behaviour as \citet{Rabinowitz:2011a} found in the high state. In describing the lightcurves of LSQ1725-64, \citet{Rabinowitz:2011a} divided its orbit into three regions. Following and refining this designation we refer to orbital phases between 0.97 and 1.03 as region one, between 0.03 and 0.45 as region two, and between 0.45 and 0.97 as region three. We have marked these regions in \autoref{fig:HighLowPhot}. Note that these numbers are adjusted from \citet{Rabinowitz:2011a} to incorporate the better time resolution of our data.

Region one is the eclipse of the white dwarf and accretion stream by the secondary star. As \citet{Rabinowitz:2011a} point out it is one of the deepest eclipses discovered, owing to the faintness of the secondary, and to the orientation of the accreting pole which makes it brightest just before eclipse ingress. A close-up of the eclipse is shown in the bottom panel of \autoref{fig:HighLowPhot}. In the high state, the ingress is dominated by the very rapid eclipse of the accretion spot on the white dwarf photosphere (e.g. \citet{ODonoghue:2006Ma}). The relatively slow decline after ingress, which lasts close to two minutes, or about half of the total eclipse time, we attribute to the secondary star slowly eclipsing an optically thick accretion stream that funnels material along the magnetic field and on to the accretion pole which is nearly perpendicular to the line of sight at this orbital phase (see below). This has been seen before by \citet{Bridge:2003Ma} in EP Dra who interpret it as an eclipse of the accretion stream and use it as a proxy for the Alfv\'{e}n radius, as we also do later. The eclipse of this stream is not detected in the low state lightcurve. The white dwarf egress lasts 23.4 $\pm$ 0.3 s, and we will use this value later to measure the white dwarf radius. We do not see the reemergence of the spot at the end of eclipse, indicating it has disappeared around the limb of the white dwarf. 

We interpret region two as the half of the orbit in which the accretion heated pole is on the far side of the white dwarf. In this region, just after egress, the high state lightcurve is brighter than the low state lightcurve, presumably because the bright accretion stream adds flux in the high state. 

We interpret region three as the half of the orbit in which the accreting pole is on the near side of the white dwarf, so that its light represents a substantial contribution to the overall flux. As seen in \autoref{sec:spectroscopic}, the main increase in flux comes from a broad inflation of the continuum peaked at around 5500 \AA. If we attribute this to cyclotron radiation, it should be beamed at 90 degrees to the magnetic field axis, and therefore brightest just after the accretion pole comes into view at the stellar limb, and again just before it disappears half an orbit later. The lightcurve conforms to this expectation.

From the orbital phases of accretion spot appearance and eclipse, we estimate the longitude of the accretion spot to be 99 $\pm$ 5 degrees. Using the inclination we calculate in \autoref{sec:parameters}, $ i = 85.6^{\circ} \pm 1.7^{\circ}$, and the phase length that the accretion spot is in view, $\Delta \phi = 0.55\ \pm\ 0.03$, we can estimate the co-latitude of the magnetic pole, $\beta$,
\begin{equation}
\tan(i) \times \tan(\beta) = - \sec(\pi \Delta \phi)
\end{equation}
(see \citet{Beuermann:1987a}). This result is highly sensitive to the inclination, but yields an estimate of $10^{\circ} < \beta < 59^{\circ}$.

\subsubsection{Photometric Variations During the Low State}
\label{sec:ellipsoidal}

\begin{figure}
    \includegraphics[width=0.5\textwidth]{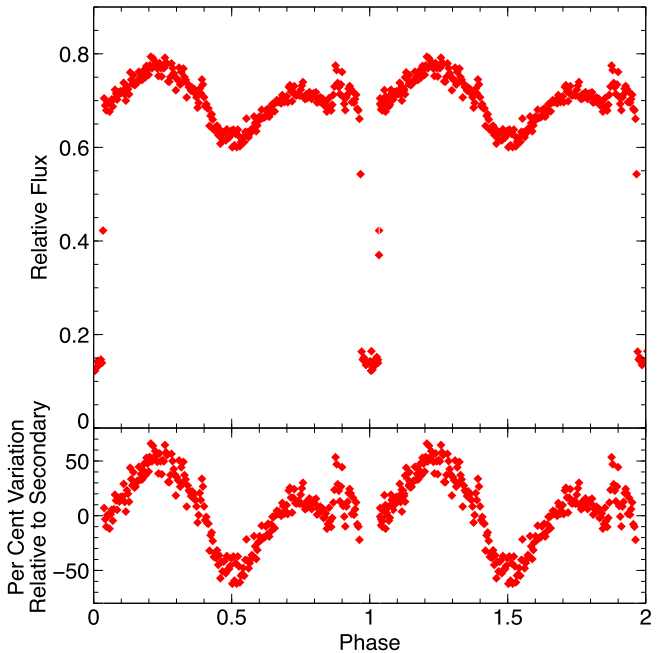}
  \caption{The low state lightcurve of LSQ1725-64 showing the amplitude variations (top). The eclipse shows the brightness of the secondary star and if the variations arise from its elliptical shape they would be at least 50 per cent of the secondary flux. We show this in the bottom panel. We hypothesize that the variations predominantly come from the two previously heated magnetic poles of the white dwarf.}
  \label{fig:LowPhot}
\end{figure}

The SOAR low-state lightcurve in white light, shown in the top panel of \autoref{fig:LowPhot}, shows a periodic modulation at twice the orbital period, with the maximum occurring near phases 0.25 and 0.75. Upon first glance, this resembles what we expect from ellipsoidal variations of the secondary, but this appearance is deceiving. As the eclipse in \autoref{fig:LowPhot} shows, in the observed band only $\sim \frac{1}{6}$ of the light comes from the secondary. After correcting for the dilution by the white dwarf, the variations would be over 50 per cent, as shown in the bottom panel of \autoref{fig:LowPhot}. We show in \autoref{app:ellipsoidal} that the largest ellipsoidal variation expected for the mass ratio in LSQ1725-64 is 11.7 per cent. Thus, these modulations cannot be caused solely by ellipsoidal variations of the secondary.

\citet{Piro:2005a} show that after dwarf novae outbursts, the white dwarf cooling time back to the quiescent temperature can be months to years. Therefore, we expect any magnetic pole on the white dwarf that has accreted in the past to be hotter than the surrounding white dwarf photosphere. We hypothesize that the low state lightcurve modulation arises from the white dwarf, and is the fluctuation of two previously heated magnetic poles of the white dwarf. Previous observations of polars have detected accretion switching between one and two poles in both QS Tel \citep{Rosen:1996Ma} and MT Dra \citep{Schwarz:2002a}. 

The two maxima in the low state align in phase with our expectation of magnetic pole locations on the white dwarf based on our description in \autoref{sec:nomenclature}, and assuming a dipole geometry. If the double-humped variation in the low state arises from the magnetic poles then the temperature of each pole might conceivably be measured, and would constrain the time-averaged accretion rate on to each pole \citep{Townsley:2009AJa}.

\subsection{Long Term Monitoring of Photometric States}
\label{sec:multiple}

\begin{figure*}
\centering
\begin{minipage}[c]{\textwidth}
\centering
  \includegraphics[scale=0.75]{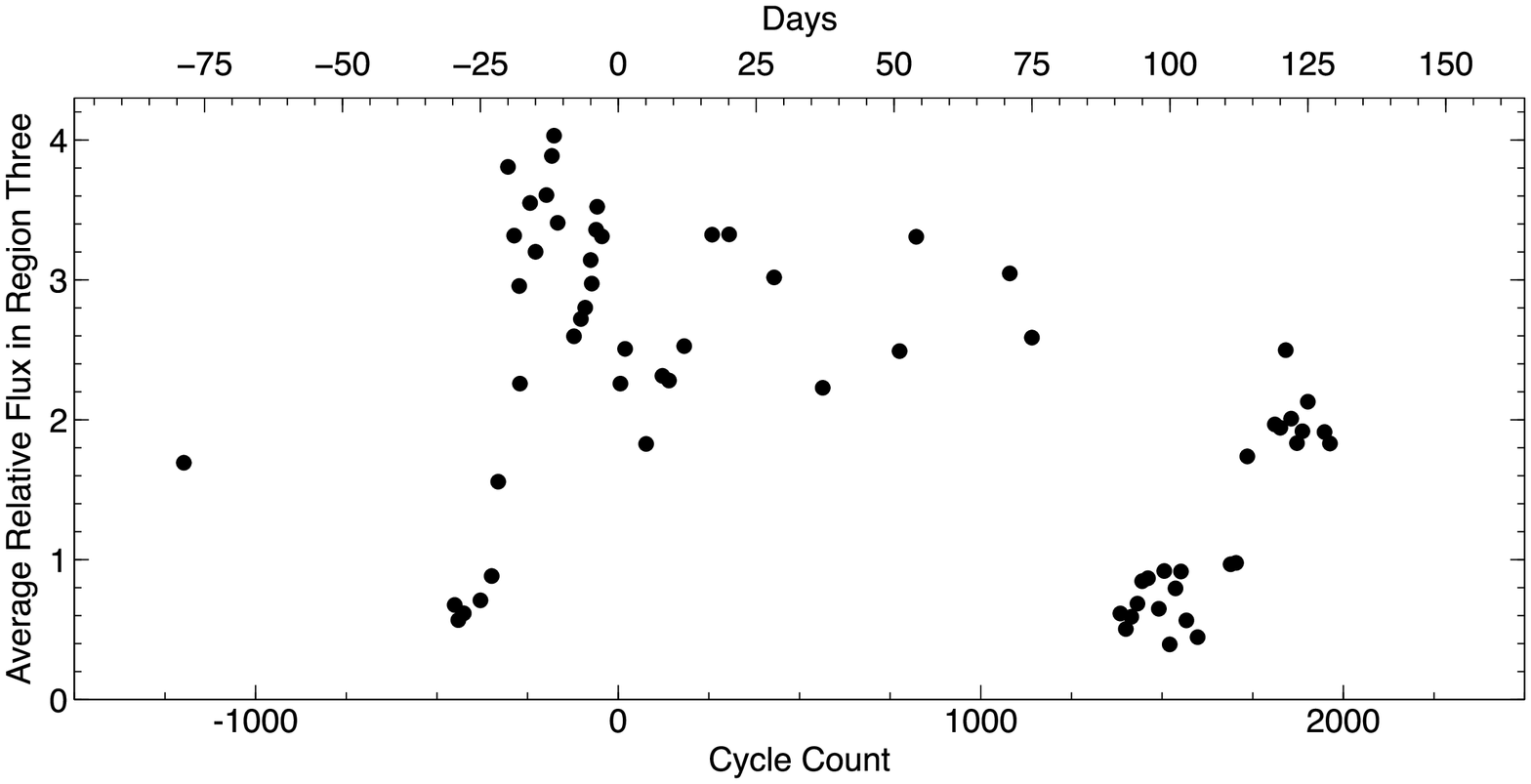}
  \caption{The average flux in region 3 from PROMPT photometry. Day 0 corresponds to the first cycle (E = 0) in the ephemeris presented above. One average flux point is shown for each night when the entirety of region 3 was sampled at least once. This average provides an indication of the accretion state of LSQ1725-64. We observed two transitions from the low state to the high state, one at cycle $\sim\ $-300 and another at $\sim\ $1700.}
  \label{fig:state}
\end{minipage}
\end{figure*}

\autoref{fig:state} shows the average region three flux for each night of PROMPT data, where, for consistency, we only include nights when the entirety of region three was sampled at least once. Between 6 June 2013 and 12 November 2013 we detected LSQ1725-64 in a high state on 45 out of 70 nights. These nights are when we observed with PROMPT at our highest cadence, and this number is likely a lower limit on the percentage of time LSQ1725-64 spends in a high state because the near-nightly monitoring was motivated by the detection of LSQ1725-64 in the low state. 

In individual low state lightcurves, we see evidence for sporadic mass transfer in the form of occasional high points coinciding with the largest peaks in the high state lightcurve. These high points do not occur in region two or in the central dip of region three so are likely related to the accretion poles.

The transition between polar states has been observed a few times in other polars (e.g. \citet{Gerke:2006Pa}), but the cause of these transitions and their time-scales are not generally well-understood. We were fortunate to observe the transition from the low state to the high state twice with PROMPT, as can be seen in \autoref{fig:state}. Based on the better-sampled of these two transitions, it appears to occur over three days, or $\sim$45 orbits. The second transition was interrupted by cloudy nights, but we can say that the change to the high state occurred within a 5 day period.

\begin{figure*}
\centering
\begin{minipage}[c]{\textwidth}
\centering
    \includegraphics[scale=0.68]{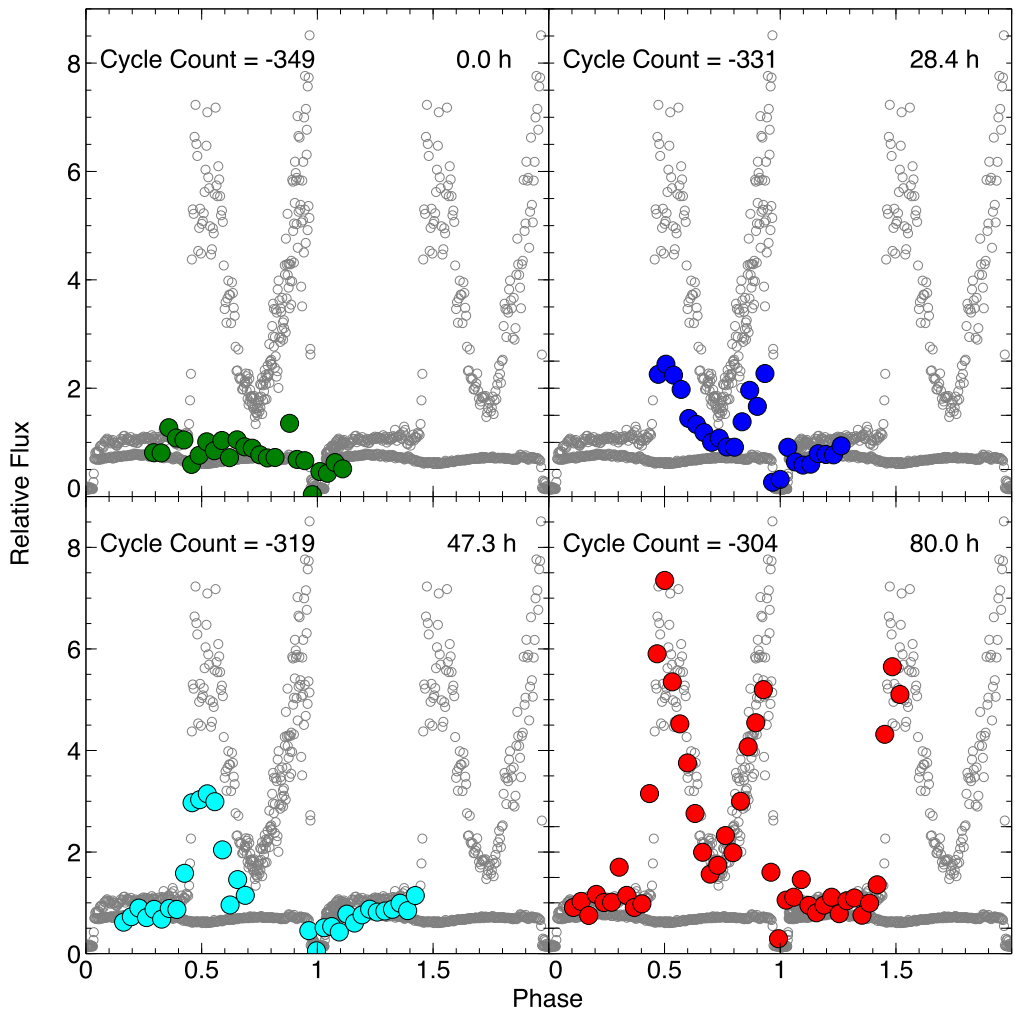}
  \caption{Nightly PROMPT monitoring (colours) showing the transition from the low state (top left) to the high state (bottom right). SOAR low and high state lightcurves are repeated in grey for reference. The cycle counts and times correspond to the leftmost side of each figure. We observe the return to highest state to occur over $\sim$45 orbits.}
  \label{fig:transition}
\end{minipage}
\end{figure*}

To examine this state change closely, we have compiled a sequence of light curves around the first transition, and included all the PROMPT data, even those where region three was not fully sampled. \autoref{fig:transition} shows a time sequence of these PROMPT lightcurves at the time of the first transition from the low to the high state, with cycle numbers referenced to the zero point in our ephemeris. Each panel shows the PROMPT lightcurve superposed on the SOAR high and low state lightcurves of \autoref{fig:HighLowPhot}. A monotonic increase in brightness is seen progressing after cycle -349 and reaching maximum by -304. These data support models in which the increase in mass transfer rate is a not an immediate change, but one that unfolds over days. This time-scale is similar to the $\sim$day transition times predicted by the starspot model (see \citet{King:1998a}) and does not rule out the suggestion by \citet{Wu:2008a} that the white dwarf magnetic field plays a significant role in the high and low states.


\section{Spectroscopic Properties of LSQ1725-64}
\label{sec:spectroscopic}

In this section, we will present our reduced and flux calibrated spectra organized by accretion state and orbital phase. 

\subsection{Continuum Measurements}
\label{sec:continuummeasurements}

\begin{figure*}
\centering
\begin{minipage}[c]{\textwidth}
\centering
  \includegraphics[scale=0.90]{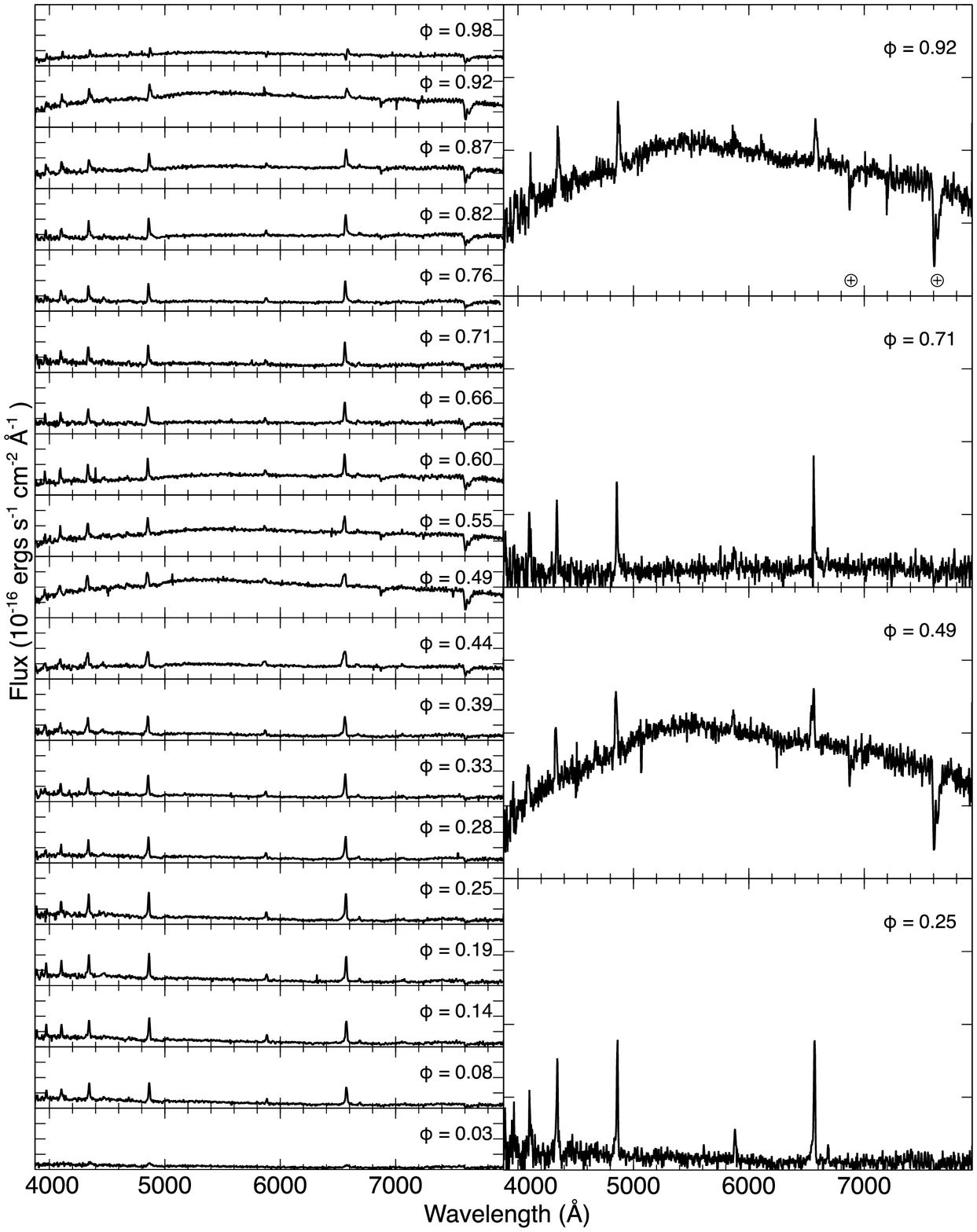}
  \caption{Left: Time-series spectroscopy of LSQ1725-64 in a high state from 2013-07-27. Time proceeds up the page. Right: Same as the left, but with low state spectra from the corresponding orbital phase subtracted. We attribute the broad maximum centred around 5500 \AA\ to a blended set of closely space cyclotron harmonics.}
  \label{fig:phasedspec}
\end{minipage}
\end{figure*}

In region 3, the continuum in the high state is broadly inflated compared to region 2, and to the low state. It does not show any distinct cyclotron humps, but has a maximum in the optical band around 5500 \AA. This can be seen in \autoref{fig:phasedspec} which shows a time-series of LSQ1725-64 in the high state. To illustrate the \textit{increase} in the continuum during the high state, the panels on the right show high state spectra with the low state spectra of corresponding phase subtracted. Those labelled $\phi$ = 0.49 and 0.92 correspond to maxima in the photometric lightcurve and in the spectroscopy. By analogy with BL Hydri \citep{Schwope:1995a}, the extra light we see in the continuum is consistent with a blended set of closely spaced cyclotron harmonics that appear when the accreting pole is in view. We estimate we are seeing harmonics around n = 17 given our measured field strength of 12.5 MG. \citet{Wickramasinghe:1985Ma} show that at high viewing angles to the magnetic field axis, like we have at phases 0.49 and 0.92, most of the cyclotron intensity is concentrated at higher harmonics. Polarization measurements during the high state are required to confirm this view.

\begin{figure*}
\centering
\begin{minipage}[c]{\textwidth}
\centering
    \includegraphics[scale=0.68]{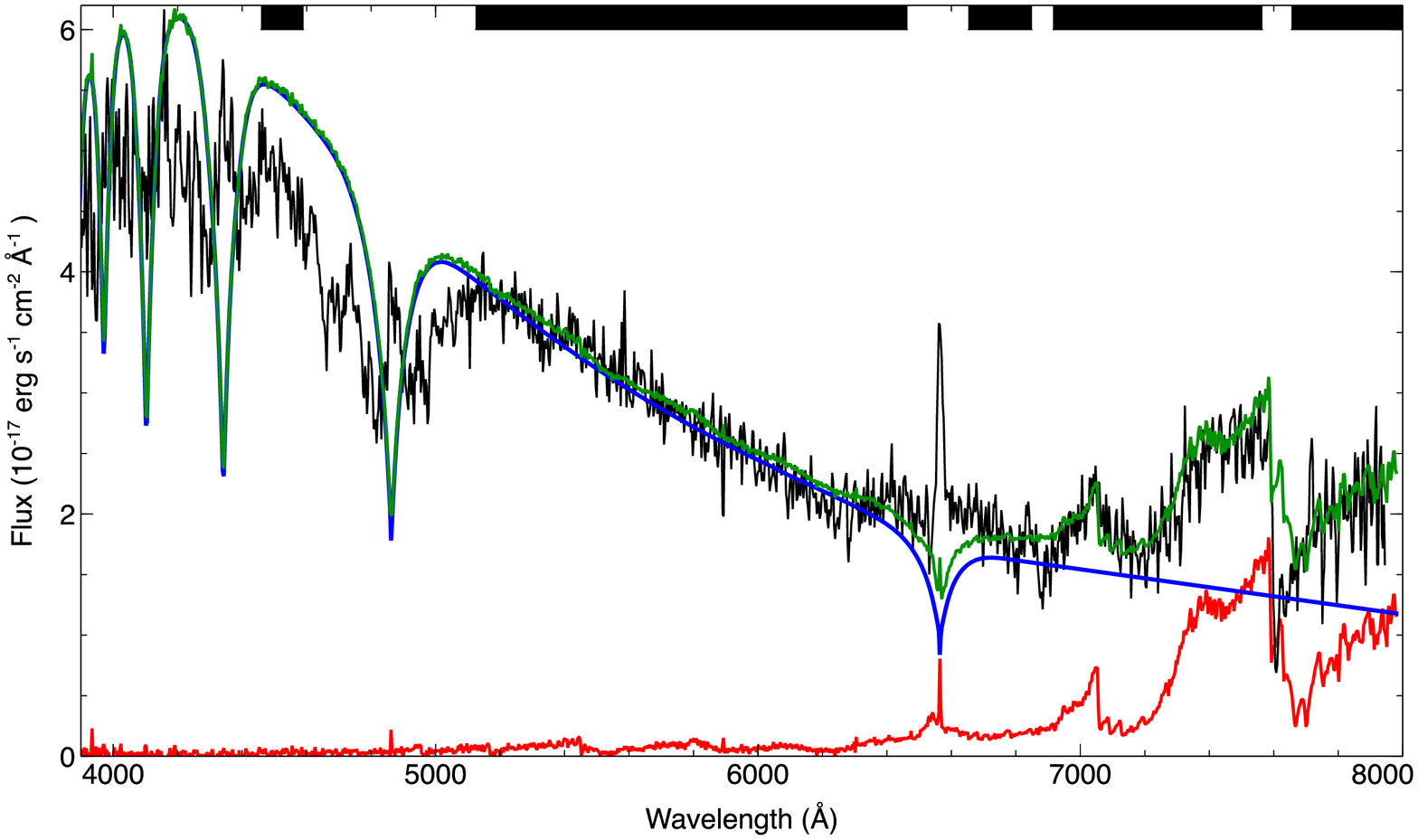}
  \caption{Combined fit to the low state orbital averaged spectrum of LSQ1725-64 (black) with an M8 (red) secondary and 12650 K and log(g) = 8.5 non-magnetic white dwarf (blue). The combined white dwarf + M8 spectrum is shown in green. The black bars at the top show the regions of the spectrum that were used in the fit. We excluded the Balmer lines, emission lines, and telluric absorption.}
  \label{fig:components}
\end{minipage}
\end{figure*}

In the low state, regions 2 and 3 are more similar to each other, consistent with a dramatic reduction or cessation of accretion. We have averaged all the low state spectra to improve the signal-to-noise and yield a spectrum that is relatively uncontaminated by accretion light (\autoref{fig:components}). The continuum is dominated by the white dwarf photosphere on the blue end and features from the secondary star on the red end. We have excluded region 1 from our average. 

To estimate the white dwarf temperature and explore the nature of the secondary, we have fitted the average spectrum of \autoref{fig:components} with combined non-magnetic white dwarf models from \citet{Koester:2010Ma} and M-dwarf templates from \citet{Bochanski:2007a}. We used a white dwarf grid in steps of 250 K at log(g)=8.5 (see below) and an M-dwarf grid of integer spectral types. We generated ensembles of model combinations that were simple sums of white dwarf model spectra with M-dwarf model templates. The free parameters in the fits were the white dwarf model temperature, the M-dwarf spectral type, and normalization factors for each star that account for their distances. Following \citet{Rebassa-Mansergas:2007Ma}, we did not insist that the distances be the same. In the fit of the models to the data, we only included regions of the spectrum not contaminated by Zeeman splitting, emission lines, or telluric absorption. We show the included regions as black bars at the top of the figure. Because this excludes the Balmer lines, we did not attempt to fit the surface gravity of the white dwarf. Instead, we use the best radius and surface gravity derived later from eclipse parameters (\autoref{sec:parameters}) and the white dwarf mass-radius relation to estimate log(g) $\sim$ 8.5. We tested different values of log(g) and found that the white dwarf temperature is relatively insensitive to log(g). 

Our best-fitting composite model is shown in \autoref{fig:components}. The poor fit at the shortest wavelengths is due to poor flux calibration due to extinction and differential slit losses (see \autoref{sec:observations}). The best-fitting composite model has a white dwarf temperature of 12650 $\pm$ 550 K and a secondary spectral type of M8 $\pm$ 0.5. The secondary spectral type is insensitive to the uncertainty in the white dwarf temperature. The fitted distances are 326 $\pm$ 93 pc and 222 $\pm$ 112 pc for the white dwarf and M-dwarf respectively. Taking a weighted average, we estimate the best distance to LSQ1725-64 is 284 $\pm$ 71 pc.

\subsection{Emission Line Measurements}

The high state spectra show strong Balmer emission lines along with He I throughout every phase except eclipse, while in the low state the Balmer lines are reduced and He I is absent (see \autoref{fig:phasedspec} and \autoref{fig:components}). We also detect weak He II (4686 \AA) in some but not all of our spectra. In a few of our spectra on 2013-07-03, which were made in a setup with a bluer limit, we also detect Ca II at 3933 \AA. The strength varies significantly with orbital phase and is strongest around phase 0.60. This phasing of the Ca II suggests that it also comes from the irradiated photosphere of the secondary, but we do not have enough signal to noise in individual spectra to confirm using radial velocities. 

In the low state, we find that emission is less prominent (\autoref{fig:components}) and will show in \autoref{sec:rv} that it arises from a combination of accretion stream and photospheric emission from the secondary. This is not unexpected. \textit{XMM-Newton} observations of polars in low states by \citet{Ramsay:2004Ma} show that in the low state, accretion usually does not stop entirely but continues at a much reduced rate. Our modest resolution, R$\sim$600 or 500 $km\ sec^{-1}$ at H $\alpha$, means we are only able to resolve these components at extrema in the stream velocities, so detailed study of the independent components or Doppler Tomography is not possible using our data. 

The Balmer decrement can give us some idea of the temperature and density of the line emitting region, which in the high state we presume is predominantly from a section of the ballistic accretion stream (see \autoref{sec:rv}). From high state spectra of LSQ1725-64, we measure the ratios $H \alpha / H \beta = 1.05$ and $H \gamma / H \beta = 0.86$. \citet{Williams:1991a} calculates Balmer ratios in cataclysmic variables over a range of temperatures and densities. The ratios in LSQ1725-64 roughly correspond to gas with temperature of 8000 - 10000 K and $\log(N_H)$ between 12.5 and 13.5. These values are similar to those found by \citet{Gerke:2006Pa} for their measurement of the Balmer decrement in the streams of other polars.

\begin{figure}
  \includegraphics[width=0.5\textwidth]{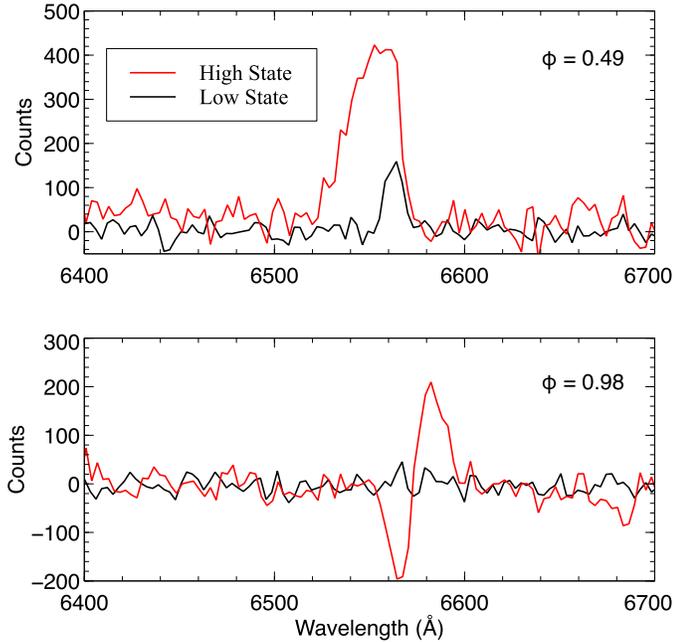}
  \caption{Comparison of H $\alpha$ emission at different phases during high and low states. The data have been continuum subtracted using the continuum flux at 6450 \AA. The accretion stream in the high state overwhelms the contribution from the secondary star near phase 0.5 (see \autoref{sec:rv}). The presence and absence of the P Cygni profile in high and low state, respectively, is shown near phase 1.}
  \label{fig:spec_Halpha}
\end{figure}

Just before eclipse, we see an absorption component appear on the blue side of the emission line, forming a P Cygni profile. This is shown in \autoref{fig:spec_Halpha}, which compares H $\alpha$ at phases 0.49 and 0.98 in the high and low state. Similar P Cygni profiles also appear in H $\beta$ and He I (5876 \AA) just before eclipse in the high state. None of these lines show absorption or P Cygni profiles in the low state. We show trailed spectra from high and low state in Online Figure 2.

\citet{Schmidt:2005a} measure P Cygni profiles in multiple lines in FL Cet just before eclipse, and following a model from \citet{Schwope:1997a} for HU Aquarii, they attribute the absorption to material in an accretion funnel just above the white dwarf. In Hu Aqr, similarly placed material causes a dip in the photometric light curve, and is described by \citet{Schwope:1997a} as being in a stagnation region where the infalling stream has attached to the magnetic field and is being redirected along field lines towards a magnetic pole. These field lines may direct the material out of the orbital plane or away from the line of centres, thereby changing the projected velocity. Near eclipse, when the projected stream velocity is at a maximum, the result will be a comparatively lower redshift, and hence the P Cygni profile we observe.

\subsubsection{Radial Velocities}
\label{sec:rv}
We measured radial velocities of $H\ \alpha,\ H\ \beta,\ H\ \gamma,$ and $He\ I\ (5876\ $ \AA) on four different nights while LSQ1725-64 was in the high state. The emission lines at each orbital phase were fitted with a gaussian. At phases where the lines could not be fitted (mostly around the eclipse) we discarded the results. We used the 5577 \AA\ sky line to adjust our wavelength solution for each spectrum. We then fitted the time series for each emission line using the equation
\begin{equation}
v_r = \gamma + K \sin \big( 2 \pi (\phi - \phi_0)\big)\ ,
\end{equation}
holding the period fixed to that measured in \autoref{sec:ephemeris} so that the fitted parameters are the systemic velocity, radial velocity amplitude, and phase. 

\begin{figure}
   \includegraphics[width=0.5\textwidth]{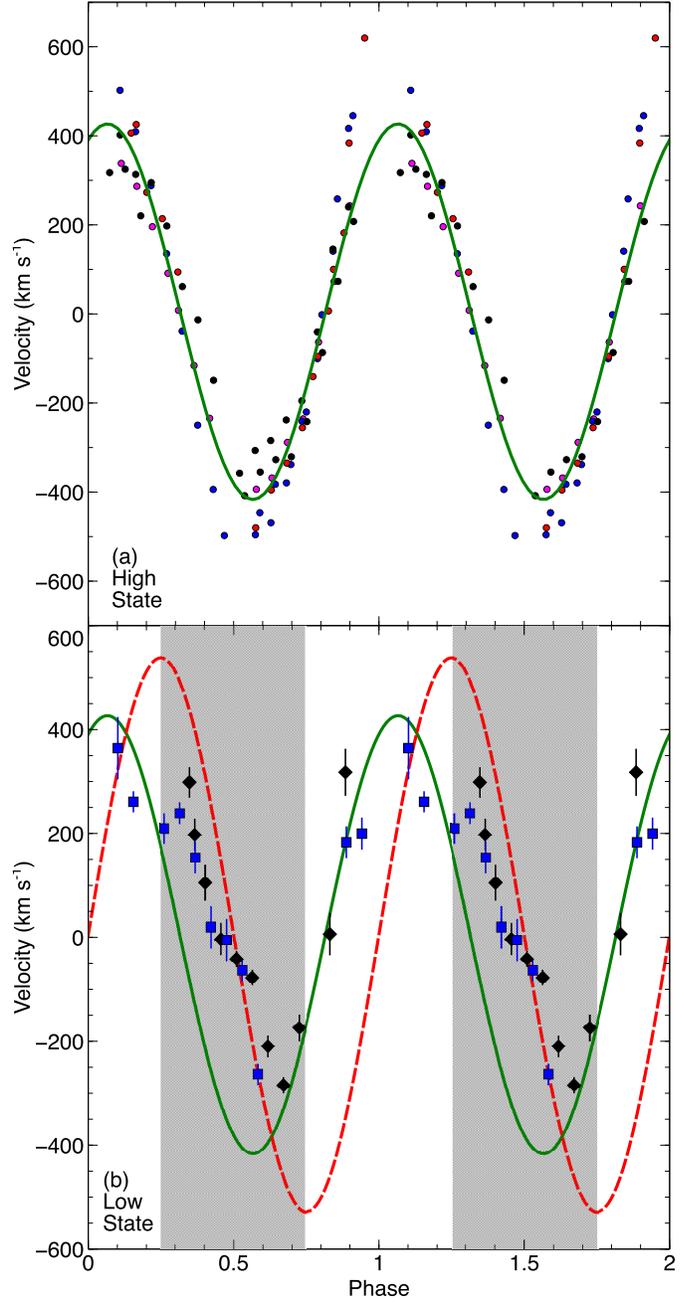}
  \caption{(a) Radial velocity measurements of H $\alpha$ in the high state. The best-fitting curve of the accretion stream velocities is shown in solid green.
(b) Radial velocity measurements of H $\alpha$ in the low state. The expected radial velocity curve of the secondary star based on our derived binary parameters in \autoref{sec:parameters} is shown in dashed red. The best-fitting curve from the high state is shown again in solid green. Different symbols and colours represent different nights. The data are plotted twice for clarity. The low state emission follows the expected accretion stream from the high state most closely when the irradiated side of the secondary is out of view (white background). When the irradiated side is in view (grey background), the data more closely follow the expected radial velocity curve of the secondary.}
  \label{fig:rv_low}
\end{figure}

We quantitatively confirm the result from \citet{Rabinowitz:2011a} that the maximum redshift occurs just after phase 0, at phase $0.044 \pm 0.001$.  This difference is roughly consistent with the expected deviation of the stream from the line of centres due to Coriolis effects \citep{Warner:1972Ma}. This reinforces the point that the primary source of emission is along the infalling ballistic accretion stream. The average radial velocity amplitude we measure is $507 \pm 18\ km\ s^{-1}$. The average systemic velocity we measure is $40 \pm 25\ km\ s^{-1}$. We show this result in the top panel of \autoref{fig:rv_low}.

We show in the bottom panel of \autoref{fig:rv_low} the fitted radial velocities of H $\alpha$ during a low state. The best-fitting H $\alpha$ curve from our high state measurements is shown as a solid green curve. The expected radial velocity from the secondary using the parameters in \autoref{sec:parameters} is shown as a dashed red curve. In phases centred on white dwarf eclipse (white background), when we see the un-irradiated side of the secondary, the data generally follow the H $\alpha$ radial velocity curve from the ballistic stream, as in the high state. However, in phases centred on 0.5 (gray background), when emission from the secondary should be at its largest, the data more closely follow the expected orbital radial velocity curve of the secondary star. 

The upper panel of \autoref{fig:spec_Halpha} compares the reprocessed H $\alpha$ in the low state with the high state emission from the stream near its maximum blueshift ($\phi$ = 0.49). Overall, the fitted radial velocities provide evidence that some of the light is from residual accretion occurring during the low state, but also that reprocessed light from the irradiated secondary is now much more easily measured. 

\section{Discussion}
\label{sec:discussion}

Our observations have refined and expanded our understanding of the properties of LSQ1725-64 as first presented by \citet{Rabinowitz:2011a}. In this section, we will discuss three key questions about LSQ1725-64 that \citet{Rabinowitz:2011a} were either unable to address or able to address only partially: the evolutionary state of the system, the masses and radii of the stars in the binary, and an estimate of the mass transfer rate.

\subsection{Is LSQ1725-64 a post-bounce system?}
\citet{Rabinowitz:2011a} suggested LSQ1725-64 is a post-bounce system based on the very red limit they derived for the \textit{V-J} colour and their fitting of binary parameters. However, their determination suffered from two limitations: their time resolution gave them a crude estimate of the eclipse length, and they did not detect the secondary during eclipse in any band other than \textit{J}. In this section we will use our better data to argue that LSQ1725-64 has properties most consistent with a pre-bounce system.

\subsubsection{Argument based on eclipse length}
\label{sec:eclipselength}

Using our highest time resolution photometry from SALT, we have measured an eclipse length from mid-ingress to mid-egress of $\Delta\phi = 0.066 \pm 0.002$, a value for the eclipse length at the upper end of the range given by \citet{Rabinowitz:2011a}.

With our more precise measurement of eclipse length, we can repeat the procedure of \citet{Rabinowitz:2011a} to solve four equations for four unknowns using a Monte Carlo method to estimate errors in the parameters. The equations are the period-mean density relationship for Roche lobe filling secondaries (\autoref{eq:pd}), the mass-radius relationship for pre- and post-bounce systems from \citet{Knigge:2006Ma}, Kepler's Law (\autoref{eq:keplerslaw}), and the relationship between inclination and eclipse length (\autoref{eq:inclination}). Following \citet{Rabinowitz:2011a} we assume in our fit a white dwarf mass of $0.75\ M_{\odot}$ and radius of $0.0013\ R_{\odot}$. 

Using a Gaussian distribution of the measured parameters, we find that only 0.08 per cent of our one million Monte Carlo simulations return a result with a large enough post-bounce secondary to create the eclipse length we measure. Therefore, we consider a post-bounce solution very unlikely. Conversely, using the pre-bounce mass-radius relation (\autoref{eq:knigge}) we find the majority of solutions have parameters close to the values we measure. 


\subsubsection{Argument based on spectral type}
\label{sec:measurementssecondary}

\begin{figure}
  \includegraphics[width=0.5\textwidth]{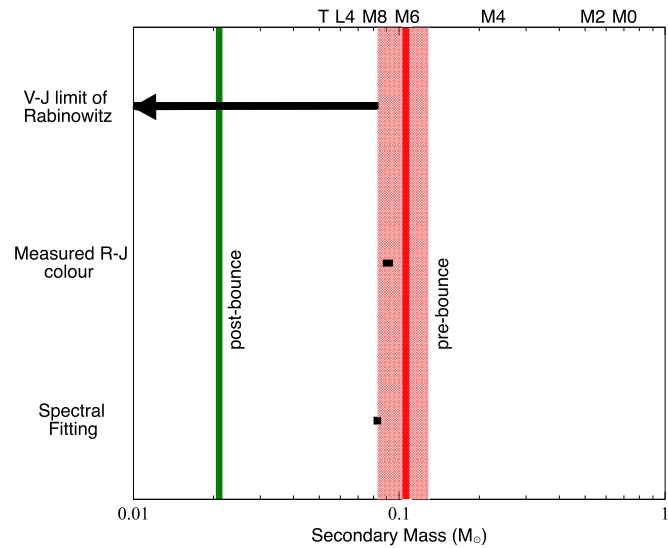}
  \caption{Constraints on the spectral type and mass of the secondary in LSQ1725-64. The green vertical line shows the post-bounce secondary mass from the models of \citet{Knigge:2006Ma}. The red vertical line and shaded region shows the pre-bounce secondary mass and distribution from \citet{Knigge:2006Ma}. We show three different constraints on the secondary mass, arising from flux measurements of the secondary. The lowest constraint is from our spectral decomposition, the middle constraint is from our \textit{R-J} colour, and the top constraint is from the \textit{V-J} limit measured by \citet{Rabinowitz:2011a}. Our results are more consistent with LSQ1725-64 being a pre-bounce system.}
  \label{fig:mdwarfs}
\end{figure}

Out best-fitting composite spectrum in \autoref{sec:continuummeasurements} yielded a spectral type of M8, with an uncertainty of about half a spectral type. This is much earlier than expected for a post-bounce system, as illustrated in \autoref{fig:mdwarfs}. The vertical bars in that figure mark the masses of secondaries on the pre-bounce (red) and post-bounce (green) sequence of \citet{Knigge:2006Ma}. Our best spectral fit lies within the scatter of the pre-bounce masses. A post-bounce solution would require a spectrum in the late T range, which is firmly ruled out by our spectrum.

The \textit{V-J} limit of \citet{Rabinowitz:2011a}, shown at the top of \autoref{fig:mdwarfs}, is just consistent with our spectral type. We also detected LSQ1725-64 in a limited number of our photometric observations in the \textit{r}$'$ and \textit{i}$'$ bands. Using the transformation of Lupton (2005)\footnote{https://www.sdss3.org/dr8/algorithms/sdssUBVRITransform.php}, we converted these into an \textit{R}-band magnitude, and combined them with the \textit{J-}band measurement of \citet{Rabinowitz:2011a} to get an \textit{R-J} colour of  $4.9 \pm 0.4$. This result also supports identification of LSQ1725-64 as a pre-bounce system.

From all of the evidence available, it appears that the secondary in LSQ1725-64 is too large, and too early in spectral type for LSQ1725-64 to be a post-bounce system.

\subsection{Binary Parameters}
\label{sec:parameters}
Because we have measured the eclipse egress length with precision, we can derive new binary parameters of LSQ1725-64 using a set of six equations instead of the four that \citet{Rabinowitz:2011a} used. We give the six equations in \autoref{sec:binary_equations}.

We solved the equations Monte Carlo style to estimate the binary parameters with errors, treating each measured value and error as a Gaussian distribution. Our inputs included the measured mid-egress time of $\phi_{egress} = 0.033 \pm  0.001$ and our measured length of the white dwarf egress of $t_{egress} = 23.4\ \pm\ 0.3\ s$, along with the orbital period. We ran one million simulations and rejected solutions that did not have secondaries large enough to match the eclipse length. Our output results and the associated $1 \sigma$ errors are shown in \autoref{tab:param}. 

\begin{table*}
\centering
\begin{minipage}{170mm}
\centering
  \caption{Binary Parameters of LSQ1725-64}
  \label{tab:param}
\begin{tabular}{c c c c c c c}
\hline
$\phi_{egress}$ & $M_{wd}\ (M_{\odot})$ & $R_{wd}\ (R_{\odot})$ & $M_{2}\ (M_{\odot})$ & $R_{2}\ (R_{\odot})$ & a $(R_{\odot})$ & $i$ (degrees) \\
\hline
$0.033 \pm 0.001$ & $0.966 \pm 0.027$ & $ 0.0085 \pm 0.0003$ & $0.114 \pm 0.021$ & $0.155 \pm 0.010$ & $0.703 \pm 0.010$ & $85.6 \pm 1.7$ \\
\hline
\end{tabular}
\end{minipage}
\end{table*}

Our white dwarf mass of $0.97 \pm 0.03\ M_{\odot}$ falls within the white dwarf mass distribution in cataclysmic variables of \citet{Zorotovic:2011a}. This value also is similar to other known white dwarf masses in polars (e.g. \citet{Cropper:1998Ma}) though the number of polars with directly measured white dwarf masses is still small. Adopting different white dwarf mass-radius relationships from \citet{Panei:2000a} for different core compositions changes our results insignificantly, except for an iron-core white dwarf, which is not something we expect. Thus LSQ1725-64 provides another indication that the white dwarfs in cataclysmic variables are significantly more massive than those of typical single white dwarfs, whose masses cluster in a narrow range around 0.63 $M_{\odot}$ \citep{Falcon:2010a}. 

\subsection{Mass Transfer Rate}
\label{sec:mdot}

Normally, the mass transfer rate in polars is measured through the accretion luminosity in the X-ray or UV combined with a distance (see e.g. \citet{King:1987Ma}). Then, if the magnetic field is known, an estimate of the Alfv\'{e}n radius, $r_{\mu}$, can also be calculated. This approach is beyond the scope of this paper, but we can make a rough estimate of the mass transfer rate from calculating the lower limit of $r_{\mu}$.

\citet{Bridge:2003Ma} used the eclipse of the accretion stream they measured in EP Dra to estimate the Alfv\'{e}n radius. We can use the eclipse of the accretion stream seen in \autoref{fig:HighLowPhot} in the same way to get an estimate of $r_{\mu}$. As with EP Dra, this stream is variable as is the mass transfer rate, but will provide an estimate for $r_{\mu}$. The stream subtends an orbital phase of $\Delta \phi = 0.025 \pm 0.002$. We converted this into a projected length using the measured accretion spot longitude of 99 degrees and a co-latitude of $\beta = 90$ to get a lower limit on the Alfv\'{e}n radius of $0.120 \pm 0.009\ R_{\odot}$.

Using \citet{Mukai:1988Ma} and \citet{Warner:1995CASa}:
\begin{equation}
\label{eq:mdot}
\dot{M}_{16} = \bigg(\frac{1.45 \times 10^{10}}{r_{\mu}}\bigg)^{11/2}\ \mu_{34}^{2}\ \sigma_9^2\ M_{wd}^{-1/2}
\end{equation}
we can convert this into an estimate of the upper limit of $\dot{M}_{16}$. The resulting upper limit is $3.8 \pm 2.1 \times 10^{15} g\ s^{-1}$ or $6.1 \pm 3.4 \times 10^{-11} M_{\odot}\ yr^{-1}$, where we have used the stream radius, $\sigma_9$, from \citet{Lubow:1975a} for the appropriate mass ratio and assumed a dipole field. 

There are a number of assumptions and simplifications that are involved in this estimate. X-ray or UV measurements of LSQ1725-64 are required to provide a measurement of the mass transfer rate. However, our estimate is within the range of mass transfer rates seen in other polars \citep{Warner:1995CASa}, and similar to those derived by \citet{Araujo-Betancor:2005a} and \citet{Townsley:2009AJa} for polars with orbital periods like LSQ1725-64.   

This can be compared with the expected $\dot{M}$ from gravitational radiation alone. From \citet{Warner:1995CASa}
\begin{equation}
\dot{M} = 2.4 \times 10^{15} \frac{M_{wd}^{2/3} P_{orb}^{-1/6}(hr)}{\Big(1-\frac{15}{19}q\Big) (1+q)^{1/3}}\ g\ s^{-1},
\end{equation}
which gives a predicted mass transfer rate of $\sim2.3 \times 10^{15} g\ s^{-1}$. Our estimated limit on the mass transfer rate is consistent with this expectation from gravitational radiation, and does not require any extra source of braking to explain. This result is consistent with the overall picture of polar evolution (e.g. \citet{Wickramasinghe:1994Ma}). 

\section{Conclusions}
\label{sec:conclusions}

We have presented new observations of the short period (94 m) eclipsing binary LSQ172554.8-643839 that confirm it is a magnetic cataclysmic variable of the AM Herculis type. Photometric monitoring with the SOAR Telescope, SALT, and PROMPT has revealed high states separated by $\sim$109 days with 65 per cent duty cycle, indicative of changes in the mass-transfer state. The transition from low to high state occurs over a three day period, or about 45 orbits. Strong emission lines present during the high state show properties consistent with emission from a bright accretion stream. The stream emission lines fade in intensity during the low state, but do not disappear, indicating that mass transfer continues throughout the accretion cycle.

Using the Zeeman splitting of the H $\beta$ absorption line of the white dwarf, we have estimated the surface-averaged magnetic field to be $12.5 \pm 0.5$ MG. We have fitted the average low state spectrum to yield an estimate of the white dwarf temperature ($12650 \pm 550\ K$) and secondary spectral type ($M8 \pm 0.5$). Based on the spectral type and the eclipse length, we have established that LSQ1725-64 is in the pre-bounce evolutionary phase.

Our time-resolved photometry and spectroscopy has permitted us to make precise estimates of system parameters, including the white dwarf radius ($0.0085 \pm 0.0003\ R_{\odot}$), which implies a mass of $0.97 \pm 0.03\ M_{\odot}$. If we identify the height of the gas above the limb as an estimate of the Alfv\'{e}n radius, we estimate an upper limit to the average mass transfer rate of $6.1 \pm 3.4 \times 10^{-11}\ M_{\odot}\ yr^{-1}$. This result is consistent with expectations from angular momentum loss via gravitational radiation alone. 

Among polars, LSQ1725-64 is an ideal laboratory in many respects. The geometry is simple and measurable, different accretion states are frequent and dramatic, and the accretion on to one pole leaves a `reference hemisphere' on the white dwarf in the high state. There have been both \textit{Swift} and \textit{XMM-Newton} observations of LSQ1725-64. Analyzing these data is beyond the scope of this paper, but investigating these observations will help measure the accretion luminosity and infer a mass transfer rate. Additional multi-wavelength studies, polarization measurements, and detailed modeling of the system should illuminate many outstanding questions about the origin and evolution of this polar, and the nature of its accretion states.

\section*{Acknowledgements}
The authors thank the anonymous referee for suggestions which improved the manuscript. The authors also thank the telescope operators, day crew, and staff at SOAR and SALT for their assistance. We also thank Dean Townsley for helpful discussions. We thank D. Koester for the use of his DA models. J.T.F., E.D., and J.C.C. acknowledge support from the National Science Foundation, under award AST-1413001. J.T.F. thanks the taxpayers of North Carolina for their support as a Teaching Assistant. Some of the observations reported in this paper were obtained with the Southern African Large Telescope (SALT) under program 2013-1-UNC-RSA-002 (PI: Fuchs). Figures were made with Veusz, a free scientific plotting package written by Jeremy Sanders. Veusz can be found at \url{http://home.gna.org/veusz/}. The white dwarf cooling models used can be found at \url{http://www.astro.umontreal.ca/~bergeron/CoolingModels}. This research was made possible through the use of the AAVSO Photometric All-SkySurvy (APASS), funded by the Robert Martin Ayers Sciences Fund.

\def\jnl@style{\it}                       
\def\mnref@jnl#1{{\jnl@style#1}}          
\def\aj{\mnref@jnl{AJ}}                   
\def\apj{\mnref@jnl{ApJ}}                 
\def\apjl{\mnref@jnl{ApJL}}               
\def\aap{\mnref@jnl{A\&A}}                
\def\mnras{\mnref@jnl{MNRAS}}             
\def\nat{\mnref@jnl{Nat.}}                
\def\iaucirc{\mnref@jnl{IAU~Circ.}}       
\def\atel{\mnref@jnl{ATel}}               
\def\iausymp{\mnref@jnl{IAU~Symp.}}       
\def\pasp{\mnref@jnl{PASP}}               
\def\araa{\mnref@jnl{ARA\&A}}             
\def\apjs{\mnref@jnl{ApJS}}               
\def\aapr{\mnref@jnl{A\&A Rev.}}          
\def\memsai{\mnref@jnl{Mem. Soc. Astron. Ital.}} 

\bibliographystyle{mn2e}

\begin{thebibliography}{}

\bibitem[\protect\citeauthoryear{{Araujo-Betancor}, {G{\"a}nsicke}, {Long},
  {Beuermann}, {de Martino}, {Sion} \& {Szkody}}{{Araujo-Betancor}
  et~al.}{2005}]{Araujo-Betancor:2005a}
{Araujo-Betancor} S.,  {G{\"a}nsicke} B.~T.,  {Long} K.~S.,  {Beuermann} K.,
  {de Martino} D.,  {Sion} E.~M.,    {Szkody} P.,  2005, \apj, 622, 589

\bibitem[\protect\citeauthoryear{{Beuermann}, {Euchner}, {Reinsch}, {Jordan} \&
  {G{\"a}nsicke}}{{Beuermann} et~al.}{2007}]{Beuermann:2007a}
{Beuermann} K.,  {Euchner} F.,  {Reinsch} K.,  {Jordan} S.,    {G{\"a}nsicke}
  B.~T.,  2007, \aap, 463, 647

\bibitem[\protect\citeauthoryear{{Beuermann}, {Thomas}, {Giommi} \&
  {Tagliaferri}}{{Beuermann} et~al.}{1987}]{Beuermann:1987a}
{Beuermann} K.,  {Thomas} H.~C.,  {Giommi} P.,    {Tagliaferri} G.,  1987,
  \aap, 175, L9

\bibitem[\protect\citeauthoryear{{Bochanski}, {West}, {Hawley} \&
  {Covey}}{{Bochanski} et~al.}{2007}]{Bochanski:2007a}
{Bochanski} J.~J.,  {West} A.~A.,  {Hawley} S.~L.,    {Covey} K.~R.,  2007,
  \aj, 133, 531

\bibitem[\protect\citeauthoryear{{Bridge}, {Cropper}, {Ramsay}, {de Bruijne},
  {Reynolds} \& {Perryman}}{{Bridge} et~al.}{2003}]{Bridge:2003Ma}
{Bridge} C.~M.,  {Cropper} M.,  {Ramsay} G.,  {de Bruijne} J.~H.~J.,
  {Reynolds} A.~P.,    {Perryman} M.~A.~C.,  2003, \mnras, 341, 863

\bibitem[\protect\citeauthoryear{{Claret} \& {Bloemen}}{{Claret} \&
  {Bloemen}}{2011}]{Claret:2011a}
{Claret} A.,  {Bloemen} S.,  2011, \aap, 529, A75

\bibitem[\protect\citeauthoryear{{Clemens}, {Crain} \& {Anderson}}{{Clemens}
  et~al.}{2004}]{Clemens:2004a}
{Clemens} J.~C.,  {Crain} J.~A.,    {Anderson} R.,  2004, in {Moorwood}
  A.~F.~M.,  {Iye} M.,  eds, Ground-based Instrumentation for Astronomy
  Vol.~5492 of Society of Photo-Optical Instrumentation Engineers (SPIE)
  Conference Series, {The Goodman spectrograph}.
pp 331--340

\bibitem[\protect\citeauthoryear{Cropper}{Cropper}{1990}]{Cropper:1990a}
Cropper M.,  1990, SSR, 54, 195

\bibitem[\protect\citeauthoryear{Cropper, Mason, Allington-Smith,
  Branduardi-Raymont, Charles, Mittaz, Mukai, Murdin \& Smale}{Cropper
  et~al.}{1989}]{Cropper:1989Ma}
Cropper M.,  Mason K.,  Allington-Smith J.,  Branduardi-Raymont G.,  Charles
  P.,  Mittaz J.,  Mukai K.,  Murdin P.,    Smale A.,  1989, MNRAS, 236, 29

\bibitem[\protect\citeauthoryear{{Cropper}, {Ramsay} \& {Wu}}{{Cropper}
  et~al.}{1998}]{Cropper:1998Ma}
{Cropper} M.,  {Ramsay} G.,    {Wu} K.,  1998, \mnras, 293, 222

\bibitem[\protect\citeauthoryear{Eastman, Siverd \& Gaudi}{Eastman
  et~al.}{2010}]{Eastman:2010Pa}
Eastman J.,  Siverd R.,    Gaudi B.,  2010, PASP, 122, 935

\bibitem[\protect\citeauthoryear{Eggleton}{Eggleton}{1983}]{Eggleton:1983a}
Eggleton P.,  1983, ApJ, 268, 368

\bibitem[\protect\citeauthoryear{{Falcon}, {Winget}, {Montgomery} \&
  {Williams}}{{Falcon} et~al.}{2010}]{Falcon:2010a}
{Falcon} R.~E.,  {Winget} D.~E.,  {Montgomery} M.~H.,    {Williams} K.~A.,
  2010, \apj, 712, 585

\bibitem[\protect\citeauthoryear{{G{\"a}nsicke}, {Dillon}, {Southworth},
  {Thorstensen}, {Rodr{\'{\i}}guez-Gil}, {Aungwerojwit}, {Marsh}, {Szkody},
  {Barros}, {Casares}, {de Martino}, {Groot}, {Hakala}, {Kolb} \&
  {Littlefair}}{{G{\"a}nsicke} et~al.}{2009}]{Gansicke:2009Ma}
{G{\"a}nsicke} B.~T.,  {Dillon} M.,  {Southworth} J.,  {Thorstensen} J.~R.,
  {Rodr{\'{\i}}guez-Gil} P.,  {Aungwerojwit} A.,  {Marsh} T.~R.,  {Szkody} P.,
  {Barros} S.~C.~C.,  {Casares} J.,  {de Martino} D.,  {Groot} P.~J.,  {Hakala}
  P.,  {Kolb} U.,    {Littlefair} S.~P.,  2009, \mnras, 397, 2170

\bibitem[\protect\citeauthoryear{{Gerke}, {Howell} \& {Walter}}{{Gerke}
  et~al.}{2006}]{Gerke:2006Pa}
{Gerke} J.~R.,  {Howell} S.~B.,    {Walter} F.~M.,  2006, \pasp, 118, 678

\bibitem[\protect\citeauthoryear{Hessman, G{\"a}nsicke \& Mattei}{Hessman
  et~al.}{2000}]{Hessman:2000a}
Hessman F.,  G{\"a}nsicke B.~T.,    Mattei J.,  2000, \aap, 361, 952

\bibitem[\protect\citeauthoryear{{Holberg} \& {Bergeron}}{{Holberg} \&
  {Bergeron}}{2006}]{Holberg:2006a}
{Holberg} J.~B.,  {Bergeron} P.,  2006, \aj, 132, 1221

\bibitem[\protect\citeauthoryear{{Horne}, {Gomer} \& {Lanning}}{{Horne}
  et~al.}{1982}]{Horne:1982a}
{Horne} K.,  {Gomer} R.~H.,    {Lanning} H.~H.,  1982, \apj, 252, 681

\bibitem[\protect\citeauthoryear{{King} \& {Cannizzo}}{{King} \&
  {Cannizzo}}{1998}]{King:1998a}
{King} A.~R.,  {Cannizzo} J.~K.,  1998, \apj, 499, 348

\bibitem[\protect\citeauthoryear{{King} \& {Watson}}{{King} \&
  {Watson}}{1987}]{King:1987Ma}
{King} A.~R.,  {Watson} M.~G.,  1987, \mnras, 227, 205

\bibitem[\protect\citeauthoryear{{Knigge}}{{Knigge}}{2006}]{Knigge:2006Ma}
{Knigge} C.,  2006, \mnras, 373, 484

\bibitem[\protect\citeauthoryear{{Koester}}{{Koester}}{2010}]{Koester:2010Ma}
{Koester} D.,  2010, \memsai, 81, 921

\bibitem[\protect\citeauthoryear{{Kopal}}{{Kopal}}{1978}]{Kopal:1978a}
{Kopal} Z.,  ed. 1978, {Dynamics of close binary systems} Vol.~68 of
  Astrophysics and Space Science Library

\bibitem[\protect\citeauthoryear{{Landstreet}}{{Landstreet}}{1980}]{Landstreet:1980a}
{Landstreet} J.~D.,  1980, \aj, 85, 611

\bibitem[\protect\citeauthoryear{{Littlefair}, {Dhillon}, {Marsh},
  {G{\"a}nsicke}, {Southworth} \& {Watson}}{{Littlefair}
  et~al.}{2006}]{Littlefair:2006Sa}
{Littlefair} S.~P.,  {Dhillon} V.~S.,  {Marsh} T.~R.,  {G{\"a}nsicke} B.~T.,
  {Southworth} J.,    {Watson} C.~A.,  2006, Science, 314, 1578

\bibitem[\protect\citeauthoryear{{Lubow} \& {Shu}}{{Lubow} \&
  {Shu}}{1975}]{Lubow:1975a}
{Lubow} S.~H.,  {Shu} F.~H.,  1975, \apj, 198, 383

\bibitem[\protect\citeauthoryear{Markwardt}{Markwardt}{2009}]{Markwardt:2009a}
Markwardt C.,  2009, in Bohlender D.,  Durand D.,   Dowler P.,  eds,
  Astronomical Data Analysis Software and Systems XVIII Vol.~411 of
  Astronomical Society of the Pacific Conference Series, Non-linear
  least-squares fitting in idl with mpfit.
p.~251

\bibitem[\protect\citeauthoryear{Morris \& Naftilan}{Morris \&
  Naftilan}{1993}]{Morris:1993a}
Morris S.,  Naftilan S.,  1993, ApJ, 419, 344

\bibitem[\protect\citeauthoryear{{Mukai}}{{Mukai}}{1988}]{Mukai:1988Ma}
{Mukai} K.,  1988, \mnras, 232, 175

\bibitem[\protect\citeauthoryear{{Nucita}, {De Paolis} \& {Ingrosso}}{{Nucita}
  et~al.}{2012}]{Nucita:2012JPCSa}
{Nucita} A.~A.,  {De Paolis} F.,    {Ingrosso} G.,  2012, Journal of Physics
  Conference Series, 354, 012013

\bibitem[\protect\citeauthoryear{O'Donoghue, Buckley, Balona, Bester, Botha,
  Brink, Carter, Charles, Christians, Ebrahim, Emmerich, Esterhuyse, Evans,
  Fourie, Fourie, Gajjar, Gordon, Gumede, de Kock, Koeslag \&
  Koorts}{O'Donoghue et~al.}{2006}]{ODonoghue:2006Ma}
O'Donoghue D.,  Buckley D. A.~H.,  Balona L.~A.,  Bester D.,  Botha L.,  Brink
  J.,  Carter D.~B.,  Charles P.~A.,  Christians A.,  Ebrahim F.,  Emmerich R.,
   Esterhuyse W.,  Evans G.~P.,  Fourie C.,  Fourie P.,  Gajjar H.,  Gordon M.,
   Gumede C.,  de Kock M.,  Koeslag A.,    Koorts W.~P.,  2006, MNRAS, 372, 151

\bibitem[\protect\citeauthoryear{{Panei}, {Althaus} \& {Benvenuto}}{{Panei}
  et~al.}{2000}]{Panei:2000a}
{Panei} J.~A.,  {Althaus} L.~G.,    {Benvenuto} O.~G.,  2000, \aap, 353, 970

\bibitem[\protect\citeauthoryear{Patterson}{Patterson}{1984}]{Patterson:1984Aa}
Patterson J.,  1984, \apjs, 54, 443

\bibitem[\protect\citeauthoryear{{Piro}, {Arras} \& {Bildsten}}{{Piro}
  et~al.}{2005}]{Piro:2005a}
{Piro} A.~L.,  {Arras} P.,    {Bildsten} L.,  2005, \apj, 628, 401

\bibitem[\protect\citeauthoryear{Rabinowitz, Tourtellotte, Rojo, Hoyer,
  Folatelli, Coppi, Baltay \& Bailyn}{Rabinowitz
  et~al.}{2011}]{Rabinowitz:2011a}
Rabinowitz D.,  Tourtellotte S.,  Rojo P.,  Hoyer S.,  Folatelli G.,  Coppi P.,
   Baltay C.,    Bailyn C.,  2011, \apj, 732, 51

\bibitem[\protect\citeauthoryear{Ramsay, Cropper, Wu, Mason, Cordova \&
  Priedhorsky}{Ramsay et~al.}{2004}]{Ramsay:2004Ma}
Ramsay G.,  Cropper M.,  Wu K.,  Mason K.,  Cordova F.,    Priedhorsky W.,
  2004, MNRAS, 350, 1373

\bibitem[\protect\citeauthoryear{{Rebassa-Mansergas}, {G{\"a}nsicke},
  {Rodr{\'{\i}}guez-Gil}, {Schreiber} \& {Koester}}{{Rebassa-Mansergas}
  et~al.}{2007}]{Rebassa-Mansergas:2007Ma}
{Rebassa-Mansergas} A.,  {G{\"a}nsicke} B.~T.,  {Rodr{\'{\i}}guez-Gil} P.,
  {Schreiber} M.~R.,    {Koester} D.,  2007, \mnras, 382, 1377

\bibitem[\protect\citeauthoryear{{Rosen}, {Mittaz}, {Buckley}, {Layden},
  {Clayton}, {McCain}, {Wynn}, {Sirk}, {Osborne} \& {Watson}}{{Rosen}
  et~al.}{1996}]{Rosen:1996Ma}
{Rosen} S.~R.,  {Mittaz} J.~P.~D.,  {Buckley} D.~A.,  {Layden} A.~C.,
  {Clayton} K.~L.,  {McCain} C.,  {Wynn} G.~A.,  {Sirk} M.~M.,  {Osborne}
  J.~P.,    {Watson} M.~G.,  1996, \mnras, 280, 1121

\bibitem[\protect\citeauthoryear{{Schmidt}, {Szkody}, {Homer}, {Smith}, {Chen},
  {Henden}, {Solheim}, {Wolfe} \& {Greimel}}{{Schmidt}
  et~al.}{2005}]{Schmidt:2005a}
{Schmidt} G.~D.,  {Szkody} P.,  {Homer} L.,  {Smith} P.~S.,  {Chen} B.,
  {Henden} A.,  {Solheim} J.-E.,  {Wolfe} M.~A.,    {Greimel} R.,  2005, \apj,
  620, 422

\bibitem[\protect\citeauthoryear{{Schwarz}, {Greiner}, {Tovmassian}, {Zharikov}
  \& {Wenzel}}{{Schwarz} et~al.}{2002}]{Schwarz:2002a}
{Schwarz} R.,  {Greiner} J.,  {Tovmassian} G.~H.,  {Zharikov} S.~V.,
  {Wenzel} W.,  2002, \aap, 392, 505

\bibitem[\protect\citeauthoryear{{Schwope}, {Beuermann} \& {Jordan}}{{Schwope}
  et~al.}{1995}]{Schwope:1995a}
{Schwope} A.~D.,  {Beuermann} K.,    {Jordan} S.,  1995, \aap, 301, 447

\bibitem[\protect\citeauthoryear{{Schwope}, {Mantel} \& {Horne}}{{Schwope}
  et~al.}{1997}]{Schwope:1997a}
{Schwope} A.~D.,  {Mantel} K.-H.,    {Horne} K.,  1997, \aap, 319, 894

\bibitem[\protect\citeauthoryear{{Stetson}}{{Stetson}}{1987}]{Stetson:1987Pa}
{Stetson} P.~B.,  1987, \pasp, 99, 191

\bibitem[\protect\citeauthoryear{{Szkody}}{{Szkody}}{1998}]{Szkody:1998a}
{Szkody} P.,  1998, in {Howell} S.,  {Kuulkers} E.,   {Woodward} C.,  eds, Wild
  Stars in the Old West Vol.~137 of Astronomical Society of the Pacific
  Conference Series, {Spectroscopy of Cataclysmic Variables: Whopping Clues
  from Wiggly Lines}.
p.~18

\bibitem[\protect\citeauthoryear{Townsley \& G{\"a}nsicke}{Townsley \&
  G{\"a}nsicke}{2009}]{Townsley:2009AJa}
Townsley D.,  G{\"a}nsicke B.~T.,  2009, The Astrophysical Journal, 693, 1007

\bibitem[\protect\citeauthoryear{{Tremblay}, {Bergeron} \&
  {Gianninas}}{{Tremblay} et~al.}{2011}]{Tremblay:2011a}
{Tremblay} P.-E.,  {Bergeron} P.,    {Gianninas} A.,  2011, \apj, 730, 128

\bibitem[\protect\citeauthoryear{Warner}{Warner}{1995}]{Warner:1995CASa}
Warner B.,  1995, Cambridge Astrophysics Series, 28

\bibitem[\protect\citeauthoryear{{Warner}}{{Warner}}{1999}]{Warner:1999a}
{Warner} B.,  1999, in {Hellier} C.,  {Mukai} K.,  eds, Annapolis Workshop on
  Magnetic Cataclysmic Variables Vol.~157 of Astronomical Society of the
  Pacific Conference Series, {Low States in Cataclysmic Variables}.
p.~63

\bibitem[\protect\citeauthoryear{{Warner} \& {Peters}}{{Warner} \&
  {Peters}}{1972}]{Warner:1972Ma}
{Warner} B.,  {Peters} W.~L.,  1972, \mnras, 160, 15

\bibitem[\protect\citeauthoryear{Wheatley \& Ramsay}{Wheatley \&
  Ramsay}{1998}]{Wheatley:1998Aa}
Wheatley P.~J.,  Ramsay G.,  1998, ASPC, 137, 446

\bibitem[\protect\citeauthoryear{{Wheatley} \& {West}}{{Wheatley} \&
  {West}}{2002}]{Wheatley:2002a}
{Wheatley} P.~J.,  {West} R.~G.,  2002, in {G{\"a}nsicke} B.~T.,  {Beuermann}
  K.,   {Reinsch} K.,  eds, The Physics of Cataclysmic Variables and Related
  Objects Vol.~261 of Astronomical Society of the Pacific Conference Series,
  {Further analysis of the X-ray eclipse of OY Car measured with XMM-Newton}.
p.~433

\bibitem[\protect\citeauthoryear{Wickramasinghe \& Ferrario}{Wickramasinghe \&
  Ferrario}{2000}]{Wickramasinghe:2000Pa}
Wickramasinghe D.~T.,  Ferrario L.,  2000, PASP, 112, 873

\bibitem[\protect\citeauthoryear{{Wickramasinghe} \&
  {Meggitt}}{{Wickramasinghe} \& {Meggitt}}{1985}]{Wickramasinghe:1985Ma}
{Wickramasinghe} D.~T.,  {Meggitt} S.~M.~A.,  1985, \mnras, 214, 605

\bibitem[\protect\citeauthoryear{{Wickramasinghe} \& {Wu}}{{Wickramasinghe} \&
  {Wu}}{1994}]{Wickramasinghe:1994Ma}
{Wickramasinghe} D.~T.,  {Wu} K.,  1994, \mnras, 266, L1

\bibitem[\protect\citeauthoryear{{Williams}}{{Williams}}{1991}]{Williams:1991a}
{Williams} G.~A.,  1991, \aj, 101, 1929

\bibitem[\protect\citeauthoryear{Wu \& Kiss}{Wu \& Kiss}{2008}]{Wu:2008a}
Wu K.,  Kiss L.,  2008, \aap, 481, 433

\bibitem[\protect\citeauthoryear{{Zorotovic}, {Schreiber} \&
  {G{\"a}nsicke}}{{Zorotovic} et~al.}{2011}]{Zorotovic:2011a}
{Zorotovic} M.,  {Schreiber} M.~R.,    {G{\"a}nsicke} B.~T.,  2011, \aap, 536,
  A42

\end{thebibliography}

\appendix
\section{Eclipse Timings}
We include all eclipse timings used to update the ephemeris in \autoref{app:timings}.
\begin{table*}
\centering
\begin{minipage}{180mm}
\centering
  \caption{Eclipse timings used to update the ephemeris.}
  \label{app:timings}
\begin{tabular}{c c c c c c}
\hline
Observation Date & Instrument & Exposure Time (s) & Mid-egress Time & Mid-egress Time & O-C (s) \\
 Start (UT) & & & ($BJD_{TDB}$) & Error ($10^{-5}$ days)& \\
\hline
2011-09-07 & SALTICAM & 15.7 & 2455812.309128 & 12.3 & 2.30\\
2012-08-12 & Goodman & 20 & 2456152.522472 & 2.5 & -3.01\\
2012-08-12 & Goodman & 20 & 2456152.588247 & 1.2 & -0.15\\
2013-06-08 & Goodman & 20 & 2456451.713065 & 10.7 & -1.19\\
2013-06-08 & Goodman & 20 & 2456451.778822 & 7.6 & 0.14\\
2013-07-03 & Goodman & 20 & 2456476.563494 & 2.3 & 3.79\\
2013-07-03 & Goodman & 20 & 2456476.629218 & 1.6 & 2.33\\
2013-07-06 & SALTICAM & 1.7 & 2456480.442182 & 5.7 & -2.51\\
2013-07-12 & Goodman & 12 & 2456486.490430 & 3.1 & -1.65\\
2013-07-12 & Goodman & 12 & 2456486.556206 & 1.6 & 1.27\\
2013-08-05 & Goodman & 12 & 2456510.486178 & 2.0 & 0.07\\
2013-08-05 & Goodman & 12 & 2456510.551916 & 2.4 & -0.28\\
2013-08-14 & Goodman & 11 & 2456519.492761 & 2.8 & -2.79\\
2013-08-14 & Goodman & 11 & 2456519.558514 & 2.1 & -1.85\\
2013-08-15 & Goodman & 12 & 2456520.478891 & 8.2 & -2.40\\
2013-09-02 & SALTICAM & 1.7 & 2456538.294926 & 1.2 & 0.02\\
2013-11-14 & Goodman & 12 & 2456610.545059 & 4.2 & -1.58\\
2014-06-30 & Goodman & 12 & 2456838.603099 & 4.2 & -0.65\\
2014-06-30 & Goodman & 12 & 2456838.668777 & 4.3 & -6.23\\
\hline
\end{tabular}
\end{minipage}
\end{table*}

\section{Ellipsoidal Modulations as a Function of Mass Ratio}
\label{app:ellipsoidal}

Here we compute the theoretical dependence of ellipsoidal variation amplitude as a function of the mass ratio, $q = \frac{M_{sec}}{M_{wd}}$.

A description of ellipsoidal variability can be found in \citet{Kopal:1978a}. This has been summarized nicely by \citet{Morris:1993a}, who expanded the periodic variations into a discrete Fourier series. The fifth term from Equation 1 of \citet{Morris:1993a} gives the expected amplitude of the ellipsoidal variation for a Roche Lobe filling star. We ignore terms that are of order $(R/a)^4$ and higher to yield fractional amplitudes as 
\begin{equation}
\frac{\Delta F}{F_{sec}} = 0.15 \frac{(15+u)(1+\tau)}{(3-u)} \frac{1}{q} \Big(\frac{R_{sec}}{a}\Big)^3 \sin^2 i ,
\end{equation}
where $u$ is the limb darkening coefficient, $\tau$ is the gravity darkening coefficient, $R_{sec}$ is the radius of the secondary star, $a$ is the orbital separation, and $i$ is the inclination of the system. Based on our characterization of the secondary in \autoref{sec:measurementssecondary} and \citet{Claret:2011a} we use $u = 0.6$ and $\tau = 0.4$. Employing our best-fitting inclination of 85.6 degrees (\autoref{sec:parameters}) this becomes
\begin{equation}
0.15 \frac{(15+u)(1+\tau)}{(3-u)} \sin^2 i\ = 1.36.
\end{equation}

We can then write the fractional amplitude as
\begin{equation}
\frac{\Delta F}{F_{sec}} = \frac{1.36}{q} \Big(\frac{R_{sec}}{a}\Big)^{3} .
\end{equation}

Furthermore, \citet{Eggleton:1983a} gives an approximation to the Roche radius as
\begin{equation}
\frac{R_{sec}}{a} = \frac{0.49 q^{2/3}}{0.6 q^{2/3} + \ln{(1 + q^{1/3})}}\  ,
\end{equation}
which is accurate to within 1 per cent of the tabulation by \citet{Kopal:1978a}. This allows us to write the expected fractional amplitude as a function of the mass ratio
\begin{equation}
\frac{\Delta F}{F_{sec}} = \frac{1.36}{q} \Big(\frac{0.49 q^{2/3}}{0.6 q^{2/3} + \ln{(1 + q^{1/3})}}\Big)^{3}\ .
\end{equation}

\begin{figure}
    \includegraphics[width=0.5\textwidth]{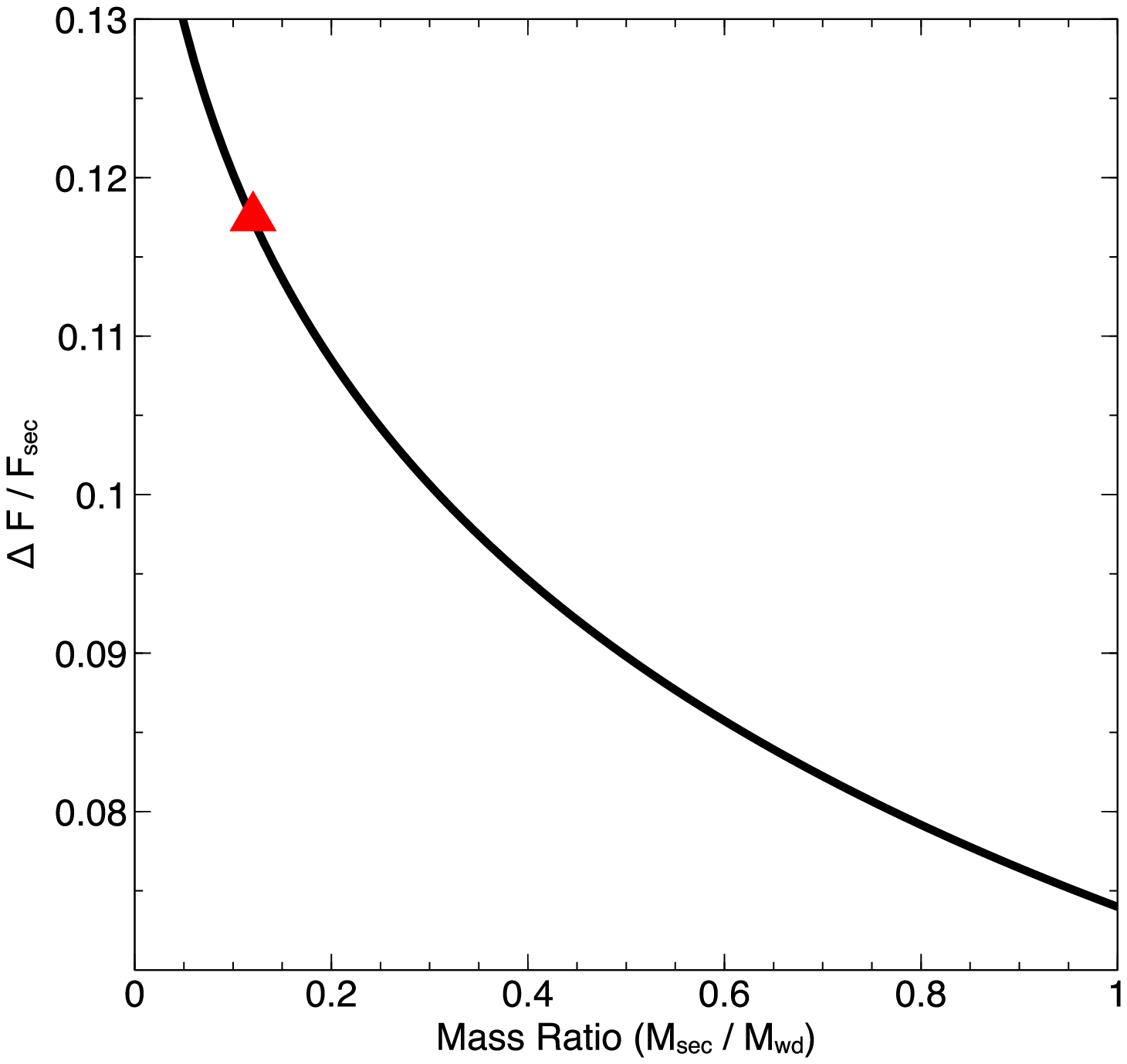}
  \caption{The expected fractional amplitude of ellipsoidal variations of a Roche lobe filling secondary as a function of mass ratio. The red triangle shows that the maximum expected fractional flux change for the mass ratio of LSQ1725-64 (\autoref{sec:parameters} is only 11.7 per cent. Thus, the 50 per cent inferred photometric variation shown in \autoref{fig:LowPhot} is too large to be explained by ellipsoidal variations alone.}
  \label{fig:MaxEllip}
\end{figure}

\autoref{fig:MaxEllip} shows this theoretical fractional amplitude caused by ellipsoidal variations as a function of mass ratio for mass ratios between 0 and 1. The red triangle shows the mass ratio ($q = 0.118$) using the masses we calculate in \autoref{sec:parameters}. The theoretical maximum of 11.7 per cent is too low to explain the measured amplitude of 50 per cent.

\section{Equations for Binary Parameter Calculation}
\label{sec:binary_equations}
The six equations we use to solve the binary parameters of LSQ1725-64 are:
\begin{enumerate}
\item The Roche condition:
\begin{equation}
\label{eq:pd}
\frac{R_2}{R_{\odot}} = 0.2361\ P_{orb, h}^{2/3} \Big(\frac{M_2}{M_{\odot}} \Big)^{1/3}\ ,
\end{equation}
where $ P_{orb, h}$ is the orbital period in hours.
\item The mass-radius relation of cataclysmic variable secondaries for the pre-bounce sequence of \citet{Knigge:2006Ma}:
\begin{equation}
\label{eq:knigge}
\frac{R_2}{R_{\odot}} = 0.230 \pm 0.008 \Big( \frac{M_2}{0.20 \pm 0.02\ M_{\odot}} \Big)^{0.64 \pm 0.02}
\end{equation}
\item The mass-radius relationship for white dwarfs from a polynomial fit to the cooling models of \citet{Holberg:2006a} and \citet{Tremblay:2011a}\footnote{see http://www.astro.umontreal.ca/~bergeron/CoolingModels/}, which assume a CO core white dwarf:
\begin{equation}
\label{eq:mrwd}
R_{wd} = -0.009297 M_{wd}^3 + 0.02669 M_{wd}^2 - 0.03678 M_{wd} + 0.02747 . 
\end{equation}

\item Kepler's Law:
\begin{equation}
\label{eq:keplerslaw}
a^3 = P_{orb}^2 \big(M_{wd} + M_2\big) \Big(\frac{G}{4 \pi^2}\Big)\ 
\end{equation} 
\item The relationship between the inclination, secondary radius, separation, and eclipse length given by \citet{Warner:1995CASa} and \citet{Horne:1982a}:
\begin{equation}
\label{eq:inclination}
\sin^2(i) \approx \frac{1- (R_2/a)^2}{\cos^2(2 \pi\ \phi_{egress})}\ ,
\end{equation}
where $\phi_{egress}$ is the phase of mid-egress and $\phi = 0$ is inferior conjunction of the secondary.
\item The length of the white dwarf egress, which depends on the path the white dwarf takes behind the secondary (see \citet{Wheatley:2002a} and \citet{Nucita:2012JPCSa}):
\begin{multline}
\frac{t_{egress}}{P_{orb}} = \\
\frac{\sqrt{[R_2 + R_{wd}]^2 - [a \cos(i)]^2} - \sqrt{[R_2 - R_{wd}]^2 - [a \cos(i)]^2}}{2 \pi a}
\end{multline}
\end{enumerate}

\end{document}